\documentclass[aps,pra,10pt,a4paper,twoside,twocolumn,superscriptaddress,amsfont,amssymb,amsmath,showpacs,preprintnumbers]{revtex4-1}

\usepackage{etoolbox}
\usepackage{amsmath}
\usepackage{bm}
\usepackage{color}
\usepackage{dcolumn}
\usepackage{float}
\usepackage{graphicx}
\usepackage{epstopdf}
\usepackage[colorlinks,linkcolor=red,citecolor=blue]{hyperref}
\usepackage{mathbbol}
\usepackage{mathrsfs}
\usepackage{placeins}
\usepackage{verbatim}
\usepackage{url}
\usepackage{breakurl}

\def\XXint#1#2#3{{\setbox0=\hbox{$#1{#2#3}{\int}$}
     \vcenter{\hbox{$#2#3$}}\kern-.5\wd0}}

\newcommand{\sgn}{\operatorname{sgn}}

\newcommand{\XUV}{{\scriptscriptstyle\mathrm{XUV}}}

\newcommand{\IR}{{\scriptscriptstyle\mathrm{IR}}}

\newcommand{\SB}{{\scriptscriptstyle\mathrm{SB}}}
\newcommand{\HH}{{\scriptscriptstyle\mathrm{H}}}
\newcommand{\RABITT}{{\footnotesize\textsc{RABITT }}}
\newcommand{\sumint}{\sum\hspace{-13pt}\int}

\begin{document}

\title{Two-photon finite-pulse model for resonant transitions in attosecond experiments}

\newcommand{\uam}{Departamento de Qu\'imica, M\'odulo 13, Universidad Aut\'onoma de Madrid, 28049 Madrid, Spain, EU}
\newcommand{\imdea}{Instituto Madrile\~no de Estudios Avanzados en Nanociencia (IMDEA-Nanociencia), Cantoblanco, 28049 Madrid, Spain, EU}
\newcommand{\ifimac}{Condensed Matter Physics Center (IFIMAC), Universidad Aut\'onoma de Madrid, 28049 Madrid, Spain, EU}
\author{\'Alvaro Jim\'enez-Gal\'an}\affiliation{\uam}
\author{Fernando Mart\'in} \email{fernando.martin@uam.es}\affiliation{\uam} \affiliation{\imdea} \affiliation{\ifimac}
\author{Luca Argenti}\email{luca.argenti@uam.es} \affiliation{\uam}

\date{\today}

\begin{abstract}
We present an analytical model capable of describing two-photon ionization of atoms with attosecond pulses in the presence of intermediate and final isolated autoionizing states. The model is based on the finite-pulse formulation of second-order time-dependent perturbation theory. It approximates the intermediate and final states with Fano's theory for resonant continua, and it depends on a small set of atomic parameters that can either be obtained from separate \emph{ab initio} calculations, or be extracted from few selected experiments. We use the model to compute the two-photon resonant photoelectron spectrum of helium below the N=2 threshold for the RABITT (Reconstruction of Attosecond Beating by Interference of Two-photon Transitions) pump-probe scheme, in which an XUV attosecond pulse train is used in association to a weak IR probe, obtaining results in quantitative agreement with those from accurate \emph{ab initio} simulations. In particular, we show that: i) Use of finite pulses results in a homogeneous red shift of the RABITT beating frequency, as well as a resonant modulation of the beating frequency in proximity of intermediate autoionizing states; ii) The phase of resonant two-photon amplitudes generally experiences a continuous excursion as a function of the intermediate detuning, with either zero or $2\pi$ overall variation.
\end{abstract}


\maketitle

\section{Introduction}

In the photoionization of poly-electronic systems, absorption of an energetic photon is often associated to the formation of autoionizing states with a lifetime of few femtoseconds. Until recently, the role of such states in photoemission could only be studied in stationary conditions, typically using the long  pulses (tens of picoseconds) generated at synchrotron facilities~\cite{Schmidt1992}. Table-top attosecond sources~\cite{Hentschel2001,Sansone2006,Goulielmakis2008,Popmintchev2010}, which deliver extreme ultraviolet (XUV) pulses capable of coherently exciting several states in the continuum across wide energy ranges~\cite{Krausz2009,Sansone2011,Chini2014}, associated to control pulses within pump-probe schemes, have opened the possibility of studying the dynamics of metastable wavepackets at its natural time scale.  For example, it has been possible to follow in time the gradual depletion of individual autoionizing states~\cite{Gilbertson2010}, as well as the progressive buildup of their population across the pump sequence~\cite{Jimenez2014}, and to reconstruct the rapid evolution of an autoionizing wavepacket~\cite{OttNature2014} from beatings between its individual metastable components.

Among attosecond interferometric spectroscopies, a prominent role is occupied by the so-called Reconstruction of Attosecond Beating by Interference of Two-photon Transitions technique (RABITT)~\cite{Paul2001b,Agostini2004}, which makes use of weak pump and probe pulses and is thus amenable to a  perturbative treatment. In RABITT spectroscopy a target atom or molecule is ionized by a train of attosecond pulses (APT), acting as a pump, in association with a weak long IR probe pulse, with a controllable time delay $\tau$ between APT and probe. The spectrum of the APT, which is generated from the interaction of an intense IR pulse with an active medium~\cite{Paul2001b,Lopez-Martens2005}, is formed by odd harmonics $\omega_{2n+1}$ of the fundamental IR frequency, $\omega_{2n+1}=(2n+1)\omega_{\mathrm{IR}}$, while the IR probe is a weak replica of the IR pulse used to generate the train. When the APT pump and the IR probe overlap, therefore, the target can either absorb one XUV photon from harmonic $2n-1$ plus one IR photon, or absorb one XUV photon from harmonic $2n+1$ and emit, in a stimulated way, one IR photon. These two processes interfere, giving rise to a sideband photoelectron signal SB$_{2n}$ which, in stationary conditions, oscillates as a function of the time delay as $I_{\SB_{2n}}=I_0\,\cos(2\omega_\IR\tau+\Delta\phi_\HH+\Delta\varphi_\mathrm{at})$~\cite{Veniard1996}, where $\Delta\phi_\HH$ is the phase difference between two consecutive harmonics in the APT spectrum,  while $\Delta\varphi_{\mathrm{at}}$, the so-called atomic phase, is the relative argument of the two-photon transition matrix elements for the IR absorption and the IR emission quantum paths.
 
If $\Delta\varphi_{\mathrm{at}}$ is a known slowly varying function of photoelectron energy, from the beating of the RABITT sidebands one can recover the relative phase between the harmonics in the train. Use of the RABITT technique with this approach has been instrumental to demonstrate that the harmonics from High-Harmonic Generation~\cite{LHuillier1993} (HHG) came in the form of trains of pulses~\cite{Paul2001b}, to understand the generation of attosecond light bursts~\cite{Dahlstrom2011} and to develop phase-compensation techniques that minimise the duration of  individual pulses within the train~\cite{Varju2005a,Mairesse2003b}. Conversely, if the harmonic phases are known, from the sideband beatings it is possible to reconstruct the atomic phases~\cite{Mauritsson2005c}. This latter approach permits one to measure both phase and amplitude of the dipole transition matrix element from the ground to the intermediate continuum states and, in turn, to reconstruct the dynamics of the photoemission event. This procedure has been followed, for example, to determine the relative delay between photoemission from the $3s$  and the $3p$ sub-shells of argon~\cite{Klunder2011a,Guenot2012}, the phase difference between photoemission from the outermost $s$ shell in different rare gases~\cite{Guenot2014,Palatchi2014}, the energy-sharing resolved double ionization of Xenon~\cite{Mansson2014}, and the nuclear dynamics in H$_2$~\cite{Kelkensberg2011}. In all these examples, the intermediate continuum states do not feature any distinct structure. The RABITT technique, therefore, could be used to infer properties of the continuum that vary smoothly across the energy gap $2\omega_\IR$ separating two consecutive sidebands in the photoelectron spectrum.

If, on the other hand, one of the two-photon paths in the RABITT scheme is nearly in resonance with a narrow intermediate bound or metastable state, both the amplitude and phase of the corresponding sidebands exhibit a strong modulation as a function of the detuning of the energy of the harmonic closest to the resonance~\cite{Johnsson2007,Ranitovic2010}.  When the contribution of the continuum to the resonant path is negligible, the sideband phaseshift undergoes a jump of $\pi$ as the frequency of the harmonics, increased gradually, traverses the energy of the intermediate resonant state. This phenomenon was observed experimentally with helium, using the $1s3p$ Rydberg state as intermediate resonance~\cite{Swoboda2010}, as well as in the N$_2$ molecule~\cite{Haessler2009a,Caillat2011}, where the intermediate resonance was an autoionizing vibronic state. Even in the latter case, the contribution of the intermediate continuum turns out to be negligible, despite the fact that the resonance in N$_2$ does interact with the ionization channel to which it eventually decays. 

In general, however, both the continuum and the localized part of an intermediate resonant state may contribute to the two-photon transition. In a recent work~\cite{Jimenez2014}, we showed that in such case, instead of undergoing a distinct jump of $\pi$, the sideband phaseshift exhibits a finite excursion.  For short pulse durations, furthermore, not only is the phase of the sideband oscillations affected by an intermediate resonant state; the frequency of the sideband beating is modified as well. For very short pulses, in fact, even in absence of an intermediate resonant state, the beating frequency is red shifted with respect to the $2\omega_\IR$ nominal value.  In the present work we provide a full derivation of an analytical model, first presented in~\cite{Jimenez2014}, which explains all these phenomena. The model provides resonant second-order ionization amplitudes, in which both the intermediate and the final continuum states of the two-photon transition matrix element can feature isolated resonances, each described with Fano's formalism, and the external field is formed by an arbitrary number of Gaussian pulses. In~\cite{Jimenez2014}, by comparing the predictions of such model with accurate \emph{ab initio} simulations~\cite{Argenti2010,ArgentiATAS2012,Argenti2014,Jimenez2014}, we showed that this approach permits us to reproduce quantitatively the ionization of the helium atom to the region of the $N=2$ autoionizing states with a sequence of attosecond pulses in association with a weak IR probe pulse. 

The paper is organised as follows.  In Sec.~\ref{sec:TwoPhoton} we give an overview of the main formulas for two-photon transition amplitudes with finite pulses, we introduce and justify the \emph{on-shell} approximation for the calculation of continuum-continuum transition matrix elements, and comment several aspects of non-resonant RABITT with finite pulses, including an explanation of the red shift of the sideband beating. In Sec.~\ref{sec:ResonantModel}~we derive the finite-pulse two-photon resonant model in the single-channel case as well as some straightforward generalisations.  In section \ref{sec:Helium} we apply the model to the resonant photoionization of helium with the RABITT technique and compare the analytical predictions with the numerical results we obtain with accurate \emph{ab-initio} numerical simulations.  In Section \ref{sec:Conclusions} we draw our conclusions.

\section{Two-photon transitions}\label{sec:TwoPhoton}
In this section we derive the lowest-order perturbative expression for  two-photon transition amplitudes with finite pulses, in both time and frequency formulation, and use it to comment on general properties of the RABITT spectroscopy in the non-resonant case. 

In dipole approximation, the total hamiltonian $\mathcal{H}(t)$ of the target atom or molecule in interaction with a light pulse is given by a field-free component $H$ plus an interaction term~\cite{Joachain}, 
\begin{equation}
\mathcal{H}(t)=H + F(t) \mathcal{O},\qquad \mathcal{O}=\hat{\epsilon}\cdot\vec{O},
\end{equation}
where $\vec{F}(t)=F(t)\hat{\epsilon}$ is the external transverse light field, which for simplicity we assume to have constant polarization $\hat{\epsilon}$, and $\vec{O}$ is an appropriate dipole operator. For example, in velocity gauge $\vec{F}(t)$ is the vector potential $\vec{A}(t)$ and $\vec{O}$  is proportional to the total canonical electron momentum $\vec{P}=\sum_{i=1}^{N_e} \vec{p}_i$, $\vec{O}=\alpha\vec{P}$, with $\alpha$ being the fine-structure constant, while in length gauge $\vec{F}(t)$ is the light electric field $\vec{E}(t)=-\alpha \partial_t \vec{A}(t)$, and $\vec{O}$ is minus the dipole moment of the system, $\vec{O}=-\vec{\mu}$.
The wavefunction $|\psi(t)\rangle$ for the system, initially in the ground state $|g\rangle$, $H_0|g\rangle=|g\rangle E_g$, is
\begin{equation}\label{eq:IntegralTDSE}
|\psi(t)\rangle=|g\rangle-i\int^t dt' F(t')\mathcal{O}_I(t')|\psi(t')\rangle,
\end{equation}
where $\mathcal{O}_I(t)=\exp(iHt)\mathcal{O}\exp(-iHt)$ is the dipole operator in the interaction picture. Unless stated otherwise, in the following we will use atomic units throughout. The r.h.s. of~\eqref{eq:IntegralTDSE} can be expanded iteratively to arbitrary order in the interaction term~\cite{Faisal},
\begin{eqnarray}
&&|\psi(t)\rangle=\sum_{n=0}^\infty |\psi^{(n)}(t)\rangle,\label{eq:PerturbativeExpansion}\\
&&|\psi^{(0)}(t)\rangle=|g\rangle,\\ 
&&|\psi^{(n+1)}(t)\rangle=-i\int^t dt' F(t')\mathcal{O}_I(t')|\psi^{(n)}(t')\rangle.
\end{eqnarray}
The question now arises as to whether a truncated version of the expansion in \eqref{eq:PerturbativeExpansion} can adequately describe attosecond pump-probe experiments. The answer depends on the intensity of the laser, its duration and the coupling strength between all the states involved. With the strong VIS and IR dressing pulses available today, the contribution of terms beyond lowest order may indeed become important~\cite{Lambropoulos1981,Bachau1986,Smirnova2003,Zhao2005,Tong2005,Zielinski2014,OttNature2014,Argenti2015b}. Rabi oscillations, for example, require the summation of the perturbative series to sufficiently high order to be reproduced across any given finite time interval. If both pump and probe ultrashort pulses are weak, however, the lowest-order approximation can be used to make accurate predictions. This is certainly the case of the \RABITT technique described in the introduction, for which the probe intensity is kept small on purpose.
The lowest perturbative transition amplitude $\mathcal{A}^{(n)}_{f g} =\langle f|\psi^{(n)}(\infty)\rangle$, from the initial ground state $|g\rangle$ to a final state $|f\rangle$, $H_0|f\rangle=|f\rangle E_f$, featuring both pump and probe contributions, appears at second order, which can be cast in the following expression,
\begin{equation}\label{eq:2PATime}
\begin{split}
\mathcal{A}_{f g}^{(2)}=-i\iint_{-\infty}^\infty dt_2dt_1 &e^{i\omega_ft_2}e^{-i\omega_gt_1}F(t_2) F(t_1)\,\,\times\\
\times&\,\,\langle f|\mathcal{O}G^+(t_2-t_1)\mathcal{O}|g\rangle,
\end{split}
\end{equation}
where we introduced the retarded Green function for the time-dependent Schr\"odinger equation of the field-free system, $G^+(t)=-i\theta(t)\exp(-iHt)$, with $\theta(x)$ being the Heaviside step function. We indicate the energy of a field-free state $|i\rangle$ indifferently as either $E_i$ or $\omega_i (=E_i/\hbar)$, and energy differences as $\omega_{ij}\equiv\omega_i-\omega_j$.
Equation~\eqref{eq:2PATime} has a well known equivalent frequency counterpart, 
\begin{equation}\label{eq:2PAFrequency}
\mathcal{A}_{f g}^{(2)}=-i\int_{-\infty}^\infty d\omega \tilde{F}(\omega_{fg}-\omega)\tilde{F}(\omega)\mathcal{M}_{f g}^{(2)}(\omega),
\end{equation}
where $\tilde{F}(\omega)= (2\pi)^{-1/2}\int F(t)\, \exp(i \omega t)\, dt$ is the Fourier Transform (FT) of the field, and $\mathcal{M}_{f g}^{(2)}(\omega)$ is a two-photon transition matrix element,
\begin{equation}
\mathcal{M}_{f g}(\omega)=\langle f | \mathcal{O} G^+(\omega_g+\omega) \mathcal{O}| g\rangle,
\end{equation}
with the retarded resolvent $G^+(\omega)$ being defined as 
\begin{equation}
G^+(\omega)\equiv\int G^+(t) e^{i\omega t} dt =(\omega-H+i0^+)^{-1}.
\end{equation}

\paragraph{The stationary regime.} 
From equation \eqref{eq:2PAFrequency}, it is easy to derive the familiar formula for stationary transition rates. To do so, 
let us suppose that the field comprises a set of overlapping square pulses $F_\alpha(t)$, with different frequencies $\omega_\alpha$ and amplitudes $F_{\alpha,0}$, but all linearly polarized along the $z$ axis and having the same duration $T$, 
\begin{equation}\label{eq:Fields}
\begin{split}
F(t) =& \sum_{\alpha} F_\alpha(t),\\
F_\alpha(t) =& F_{\alpha,0} \cos(\omega_\alpha t +\varphi_\alpha)\,\theta(T/2-|t|).
\end{split}
\end{equation}
The FT of the individual pulses can be decomposed in the sum of an absorption ($+$) and an emission ($-$) component,
\begin{eqnarray}
\tilde{F}_\alpha(\omega) &=& \tilde{F}_\alpha^+(\omega) + \tilde{F}_\alpha^-(\omega),\\
\tilde{F}_\alpha^\pm(\omega) &=& \sqrt{\frac{\pi}{2}}\,F_{\alpha,0} \,e^{\mp i\varphi_{\alpha}} \delta_T(\omega\mp\omega_\alpha),
\end{eqnarray}
where the function $\delta_T(\omega)$, proportional to the FT of the characteristic function of the $[-T/2,T/2]$ time interval, 
\begin{equation}\label{eq:deltaTasFT}
\delta_T(\omega)=\frac{1}{\sqrt{2\pi}}\mathcal{F}\left[\theta(T/2-|t|)\right](\omega)=\frac{\sin(\omega T/2)}{\pi\omega},
\end{equation}
is a representation of the Dirac delta function.
When replacing expressions~(\ref{eq:Fields}-\ref{eq:deltaTasFT}) in~\eqref{eq:2PAFrequency}, there are a limited number of contributions for which the frequency components from the two convoluted field FT overlap and which thus need to be considered,
\begin{eqnarray}\label{eq:2PAdeltaTintegral}
\mathcal{A}_{f i}^{(2)}&=&\frac{\pi}{2i}\sum_{\alpha\sigma}F_{\alpha,0} \,e^{ i\sigma\varphi_{\alpha}} \sum_{\beta\sigma'}
F_{\beta,0} \,e^{ i\sigma'\varphi_{\beta}} \,\,\times\\
&\times&\int_{-\infty}^\infty \hspace{-12pt}d\omega\,
\delta_T(\omega_{fi}+\sigma'\omega_\beta-\omega)\,
\delta_T(\omega+\sigma\omega_\alpha)\,\mathcal{M}_{f i}(\omega),\nonumber
\end{eqnarray}
where $\sigma=\mp 1$ stands for photon absorption and emission, respectively.
The last integral becomes negligible as soon as the energy-preserving condition is not satisfied, $|\omega_{fi}+\sigma\omega_\alpha+\sigma'\omega_\beta|\gg 1/T$ .  
If the two-photon matrix element $\mathcal{M}_{f i}(\omega)$ is almost constant for $|\omega+\sigma\omega_\alpha|\leq 1/T$, then we can replace it with the constant term $\mathcal{M}_{f i}(-\sigma\omega_\alpha)$ and move it out of the integral. Using the convolution theorem, $\int \tilde{f}(x-\omega)\tilde{g}(\omega)d\omega=\sqrt{2\pi}\,\widetilde{fg}(x)$, together with Eq.~\eqref{eq:deltaTasFT}, the remaining integral can be evaluated as
\begin{equation}
\int d\omega\,\delta_T(\Delta-\omega)\,\delta_T(\omega)=\delta_T(\Delta).
\end{equation}
The transition amplitudes, therefore, becomes
\begin{eqnarray}
\mathcal{A}_{f i}^{(2)}&\simeq&\frac{\pi}{2i}\sum_{\alpha\sigma}F_{\alpha,0} \,e^{ i\sigma\varphi_{\alpha}} \sum_{\beta\sigma'}
F_{\beta,0} \,e^{ i\sigma'\varphi_{\beta}} \,\mathcal{M}_{f i}(-\sigma\omega_\alpha)\,\,\times\nonumber\\
&\times&\delta_T(\omega_{fi}+\sigma\omega_\alpha+\sigma'\omega_\beta),\label{eq:2PAStationary}
\end{eqnarray}
Eq.~\eqref{eq:2PAStationary} is the familiar stationary formula expressing the transition amplitude as a coherent sum of contributions from individual time-ordered Feynman diagrams. When Eq.~\eqref{eq:2PAStationary} is valid, it is possible to define a transition rate $\mathcal{W}_{f i}=\lim_{T\to\infty} |\mathcal{A}_{f i}^{(2)}|^2/T$ on account of being  $\lim_{T\to\infty}2\pi\,\delta^2_T(\omega)/T\to \delta(\omega)$.

Equation \eqref{eq:2PAStationary} differ from \eqref{eq:2PATime} and \eqref{eq:2PAFrequency} in a fundamental way.
While either expression \eqref{eq:2PATime} or \eqref{eq:2PAFrequency} are applicable in the presence of intermediate resonant states,
the stationary expression \eqref{eq:2PAStationary} generally is not. The reason is that, the closer one gets to the resonance, the longer the exposure time required to  legitimately factor out the two-photon matrix element from the integral in \eqref{eq:2PAdeltaTintegral}. 
Thus, for pulses comparable to or shorter than the characteristic lifetime of the resonance, equation~\eqref{eq:2PAStationary} is not applicable as such, even if the truncated perturbative expression is valid. In this latter case, despite the transition being second order, a stationary regime is never achieved and a transition rate cannot consequently even be defined. Furthermore, for long exposures the second-order transition amplitude may become so large that higher-order terms, possibly infinitely many of them, are required to achieve a physically meaningful result.

\begin{figure}[hbtp!]
\begin{center}
\includegraphics[width=\linewidth]{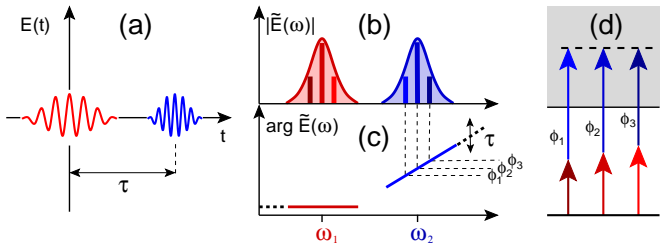}
\caption{\label{fig:TwoPhotonScheme} (Color online) Pump-probe scheme. (a) Temporal perspective: the second pulse is centered at a time $\tau$ from the first pulse, which defines the time origin. (b) With non-overlapping pump and probe spectra, the photon distribution of the pump-probe sequence does not depend on the delay between the two pulses. (c) The relative phase between different frequency components of the field, however, does depend on the time delay. (d) Since the same final energy can be reached with different combinations of the energies contained in the two pulses, the corresponding amplitudes can interfere constructively or destructively depending on their mutual phases and, in turn, on the pump-probe time delay.}
\end{center}
\end{figure}

\paragraph{The pump-probe scheme.} 
Let us now examine how a finite-pulse formulation of the second-order transition amplitude, such as Eq.~\eqref{eq:2PAFrequency}, can be used to describe a pump-probe process. In pump-probe experiments the total external field is the sum of a pump field $F_1(t)$, which can be assumed not to depend on the pump-probe time delay, thus defining the time reference, and of a probe field $F_2(t;\tau)\equiv F_2(t-\tau)$ delayed with respect to the pump by a time lapse $\tau$ (see Fig.~\ref{fig:TwoPhotonScheme}a),
\begin{equation}
F(t)=F_1(t) + F_2(t-\tau).
\end{equation}
The FT of the total field has a simple parametrization in terms of the FT of the pump pulse and of the probe pulses at zero time delay (Fig.~\ref{fig:TwoPhotonScheme}b,c),
\begin{equation}\label{eq:FTPP}
\tilde{F}(\omega)=\tilde{F}_1(\omega)+\tilde{F}_2(\omega)e^{i\omega\tau}.
\end{equation}
In a two-photon transition with finite pulses, the energy preserving condition $\omega_1+\omega_2=\omega_{fg}$ is satisfied by several different pairs of frequency components $(\omega_1,\omega_2)$, which result in separate contributions that interfere to give rise to the total transition amplitude (Fig.~\ref{fig:TwoPhotonScheme}d). Changing the time delay between pump and probe pulses alters the relative phase between all these different contributions, thus affecting the total amplitude, which becomes a function of $\tau$.

If we consider separately the positive and negative frequency components of the field ($1$ and $\bar{1}$, respectively, for the pump, $2$ and $\bar{2}$ for the probe), replacement of \eqref{eq:FTPP} in \eqref{eq:2PAFrequency} gives rise to sixteen terms associated to all possible time-ordered two-photon transitions: $21$, absorption of a pump photon followed by the absorption of a probe photon; $\bar{2}1$, absorption of a pump photon followed by the emission of a probe photon; $12$, absorption of a probe photon followed by the absorption of a pump photon, and so on. For example, the total transition amplitude for the absorption of one pump and one probe photon comprises two terms,
\begin{eqnarray}
\mathcal{A}_{f g}&=&\mathcal{A}_{f g}^{12}+\mathcal{A}_{f g}^{21},\label{eq:PPAbs}\\
\mathcal{A}_{f g}^{21}&=&-i\int_0^\infty \hspace{-10pt}d\omega \,\tilde{F}_2(\omega_{fg}-\omega;\tau)\tilde{F}_1(\omega)\mathcal{M}_{f g}(\omega),\label{eq:foldingAbs1PlusAbs2}\\
\mathcal{A}_{f g}^{12}&=&-i\int_0^\infty \hspace{-10pt}d\omega \,\tilde{F}_1(\omega_{fg}-\omega)\tilde{F}_2(\omega;\tau)\mathcal{M}_{f g}(\omega),
\end{eqnarray}
which correspond to the time-ordered diagrams where the pump photon is absorbed first and last, respectively.  Let us consider the first case in more detail. We can expand the resolvent $G^+(\omega_g+\omega)$ in the two-photon matrix element in terms of the generalized eigenstates $|\psi_{\alpha\varepsilon}\rangle$ of the field-free system, $H|\psi_{\alpha\varepsilon}\rangle=|\psi_{\alpha\varepsilon}\rangle\varepsilon$, where $\alpha$ is a collective set of quantum numbers, other than the energy, sufficient to identify any given state (channel index),
\begin{equation}\label{eq:AmpResolvent}
\mathcal{M}_{fg}(\omega)=\sum_\alpha\sumint d\varepsilon\,\,\frac{\mathcal{O}_{f,\alpha\varepsilon}\mathcal{O}_{\alpha\varepsilon,g}}{\omega_g+\omega-\varepsilon+i0^+}.
\end{equation}
If $|f\rangle$ is either a discrete state or a generalised state belonging to a featureless continuum (far from thresholds and from any resonant state), and the intermediate states contributing to~\eqref{eq:AmpResolvent} are either similarly featureless continua or discrete states far from the resonance condition (virtual excitations), then $\mathcal{M}_{f g}(\omega)$ is a smooth function of $\omega$ and, for sharply peaked field spectra, one recovers the familiar quasi-stationary expression for $\mathcal{A}_{f g}^{(2)}$ as a finite sum of weighted Feynman diagrams. In the presence of intermediate resonant states with lifetime comparable to or longer than the duration of the light pulses involved, however, $\mathcal{M}_{f g}(\omega)$ has a sharp dependence on $\omega$ and the transition never achieves a stationary regime. In this latter case, the folding with the field in Eq.\eqref{eq:2PAFrequency} must be evaluated to the full. 

\paragraph{The \emph{on-shell} approximation.}
It is worth examining the special case for Eq.~\eqref{eq:AmpResolvent} in which both the intermediate states $|\alpha \epsilon\rangle$ and the final state $|f\rangle=|\beta E\rangle$ are elastic-scattering featureless continuum states corresponding to a same parent ion. In this case, the largest contribution to the two-photon transition amplitude comes from the intermediate states that are degenerate or almost degenerate with the final state. This circumstance is evident if the continuum states are approximated with plane waves, which is a common assumption for energetic photoelectrons in multiphoton transitions (this approximation is employed in disguise, for example, in the strong-field~\cite{Keldysh1965,Faisal1973,Reiss1980}  and in the soft-photon~\cite{Maquet2007,Jimenez2013} approximations, both of which are known to work well sufficiently above threshold). Indeed, since plane waves are eigenstates of the dipole operator in velocity gauge, the only non-vanishing dipole transition matrix element is the one between two identical plane waves, 
\begin{equation}
\langle \vec{k}|\,\hat{\vec{p}}\,|\vec{k}'\rangle=\vec{k}\,\delta^{(3)}(\vec{k}-\vec{k}')=\vec{k}\delta^{(2)}(\hat{k}-\hat{k}')\frac{\delta(E-E')}{\sqrt{2E}}.
\end{equation}
Notice that such approximation applies when estimating the two-photon transition matrix element from a \emph{bound state} $|g\rangle$ to the continuum,
\begin{equation}
\langle \vec{k}|p_z G^+(E_g+\omega) p_z|g\rangle \simeq \frac{k_z\,\langle \vec{k}| p_z|g\rangle}{E_g+\omega-{k}^2/2+i0^+}.
\end{equation} 
It does not imply, however, any net absorption or emission of photons between free-electron states, which is and remains a prohibited process. We will call \emph{on-shell} approximation the assumption that the transition matrix element between unstructured continuum states is diagonal in the energy, 
\begin{equation}\label{eq:OnShellApproximation}
\langle \beta E'|\mathcal{O}|\alpha E\rangle\simeq \bar{O}_{\beta\alpha}(E) \delta(E-E'),
\end{equation}
where $\bar{O}_{\alpha\beta}(E)$ is the integral of the actual transition amplitude $\mathcal{O}_{\beta E,\alpha \epsilon}$ in a narrow energy interval $\epsilon\in (E-\delta,E+\delta)$ to which significant values of $\bar{O}_{\alpha\beta}(E)$ are hopefully restricted,
\begin{equation}
\bar{O}_{\beta\alpha}(E)=\int_{E-\delta}^{E+\delta} \langle\beta E|\mathcal{O}|\alpha\varepsilon\rangle\,d\varepsilon.
\end{equation}

The \emph{on-shell} approximation is quite acceptable even when considering radiative transitions between the Coulomb or shifted Coulomb waves commonly encountered in atomic ionization, and it becomes increasingly more accurate as the electron energy increases. 
\begin{figure}[hbtp!]
\begin{center}
\includegraphics[width=\linewidth]{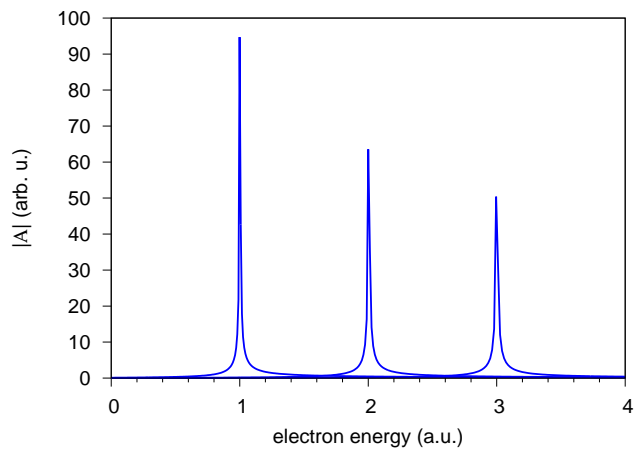}
\caption{\label{fig:ContContTransAmp} 
Absolute value, in atomic units, of the exact analytical reduced velocity-gauge dipole matrix element $\langle\psi_{E_p}\|\mathcal{O}^{v}_{1}\|\psi_{E_s}\rangle$ of the hydrogen atom, from three selected $s$ scattering states ($E_s=1,\,2,\,3$~a.u.), to several $p$ states in the continuum. The sharp localisation of the amplitude at $E_p\simeq E_s$ underpins the validity of the \emph{on-shell} approximation. For more details, see~\cite{Marante2014}.
}
\end{center}
\end{figure}
For example, Fig.~\ref{fig:ContContTransAmp} (see also \cite{Marante2014}) shows the continuum-continuum transition matrix elements in the hydrogen atom from three selected initial scattering states with $\ell=0$ and energies $E_s=$1, 2, 3~a.u., to $\ell=1$ scattering states as a function of the energy $E_p$ of the final states. It is clear that the transition amplitudes are strongly peaked at $E_p=E_s$. In conclusion, using the \emph{on-shell} approximation, the non-resonant two-photon transition matrix element from an initial state $|g\rangle$  to a final continuum state $|\beta E\rangle$ through intermediate continuum states $|\alpha\varepsilon\rangle$, $\mathcal{M}^{(\alpha)}_{\beta E,g}(\omega)=\langle \beta E | \mathcal {O} G^+(E_g+\omega)Q_\alpha\mathcal{O}|g\rangle$, where $Q_\alpha$ is the projector on the intermediate continuum $\alpha$, can be written as
\begin{equation}\label{eq:MPolarStructure}
\mathcal{M}^{(\alpha)}_{\beta E,g}(\omega)\simeq \frac{\bar{\mathcal{O}}_{\beta\alpha}(E)\mathcal{O}_{\alpha E,g}}{E_g+\omega-E+i0^+}.
\end{equation}
The two-photon transition matrix element $\mathcal{M}_{fg}(\omega)$ has thus assumed the form of a rational function which, apart for a simple pole in the lower half of the complex plane, depends only weakly on the frequency $\omega$. In the next section we will see that, with some additional approximations, $\mathcal{M}_{fg}(\omega)$ can be cast in a form similar to Eq.~\eqref{eq:MPolarStructure} even in the presence of intermediate and final resonance states. When this is the case, folding with the field components, as in Eq.~\eqref{eq:foldingAbs1PlusAbs2}, can be computed analytically for certain shapes of the light pulses. In the following, we will examine the relevant case of Gaussian pulses. We will subsequently apply the formula to the case of the non-resonant RABBIT transition and examine the effect of finite pulse duration on the RABITT beating frequency. The more general case of intermediate and final resonant states will be treated in Sec.~\ref{sec:ResonantModel}.

\paragraph{Gaussian pulses.}\label{sec:IId} The vector potential of an ultrashort laser pulse can be conveniently approximated with a linearly polarised Gaussian pulse $\vec{A}(t)=\hat{z}A(t)$ parametrized as
\begin{equation}\label{eq:GaussianPulse}
A(t)= A_0 \mathrm{e}^{-\frac{\sigma^2}{2}(t-t_0)^2} \cos [\omega_0 (t-t_0)+\varphi],
\end{equation}
where $A_0$, $\omega_0$, $t_0$, $\sigma$ and $\phi$ are the amplitude, carrier angular frequency, central time, spectral width and carrier-envelope phase of the pulse, respectively. Several Gaussian pulses can be combined to give rise to arbitrary pulse sequences, or to chirped pulses.
The absorption and emission components in the FT of a single Gaussian pulse, $\tilde{A}(\omega)=\tilde{A}^+(\omega) + \tilde{A}^-(\omega)$,  are 
\begin{equation}\label{eq:FourierTransformPulse}
\tilde{A}^{\pm}(\omega) = \frac{A_0}{2\sigma}\, \mathrm{e}^{i (\omega t_0 \mp \varphi)} \mathrm{e}^{-\frac{(\omega \mp \omega_0)^2 }{2\sigma^2}}.
\end{equation}
As shown in App.~\ref{app:Faddeeva}, in the case of Gaussian pulses, the folding of the field with a simple-pole function $(\omega-z_0)^{-1}$, as in Eq.~\eqref{eq:foldingAbs1PlusAbs2}, can be expressed in closed form,
\begin{equation}
\int_{-\infty}^\infty \hspace{-10pt}d\omega \,\frac{\tilde{F}_2(\omega_{\beta E, g}-\omega;\tau)\tilde{F}_1(\omega)}{E_g+\omega-E+i0^+}=i\mathcal{F}^{21}(\tau)\,e^{i\omega_2\tau}\,w(z_E^{21}),\label{eq:frequencyIntegral}
\end{equation}
where $\mathcal{F}^{21}(\tau)$ is a form factor of the pulse sequence
\begin{eqnarray}
\mathcal{F}^{21}(\tau)&=&-\pi\frac{A_{1}A_{2}}{4\sigma_1\sigma_2}\,\mathrm{e}^{-i(\varphi_1+\varphi_2)}\times\nonumber\\
&\times&\exp\left[-\frac{1}{2}\left(\frac{\delta^2}{\sigma^2}+\frac{\tau^2}{\sigma_t^2}+2i\frac{\sigma_2}{\sigma_1}\frac{\delta}{\sigma}\frac{\tau}{\sigma_t}\right)\right],
\end{eqnarray}
with $\sigma=\sqrt{\sigma_1^2+\sigma_2^2}$, $\sigma_t=\sqrt{\sigma_1^{-2}+\sigma_2^{-2}}$, and $\delta=E_g+\omega_1+\omega_2-E$,
while the complex parameter $z_E^{21}$ is defined as
\begin{equation}
z^{21}_E=\frac{\sigma_t}{\sqrt{2}}\left[\left(\omega_1-\frac{\sigma_1^2}{\sigma^2}\delta-i\frac{\tau}{\sigma_t^2}
\right)-E+E_g\right],
\end{equation}
with $w(z)=e^{-z^{2}}\mathrm{erfc}(-iz)$ being the Faddeeva special function.  The transition amplitude $\mathcal{A}_{\beta E,g}^{21}$, therefore, takes on the form
\begin{equation}~\label{eq:NonResonantFiniteAmplitude}
\mathcal{A}_{\beta E, g}^{21}=\mathcal{F}^{21}(\tau)\,e^{i\omega_2\tau}\,\sum_\alpha\bar{\mathcal{O}}_{\beta \alpha}(E)\mathcal{O}_{\alpha E,i}\,w(z^{21}_E).
\end{equation}
This last equation is essentially equivalent to the one formulated by Ishikawa and Ueda in terms of the Dawson integral (compare with Eq.~2 in~\cite{Ishikawa2012}).

In the region where the pulses do not overlap, the two-photon transition amplitude vanishes. How it gradually decays as a function of the pump-probe time delay is dictated by the product $\mathcal{F}^{21}(\tau)w(z^{21}_E)$, which falls off like a Gaussian for $|\tau| \gg \sigma_t$.
\begin{figure}[hbtp!]
\begin{center}
\includegraphics[width=\linewidth]{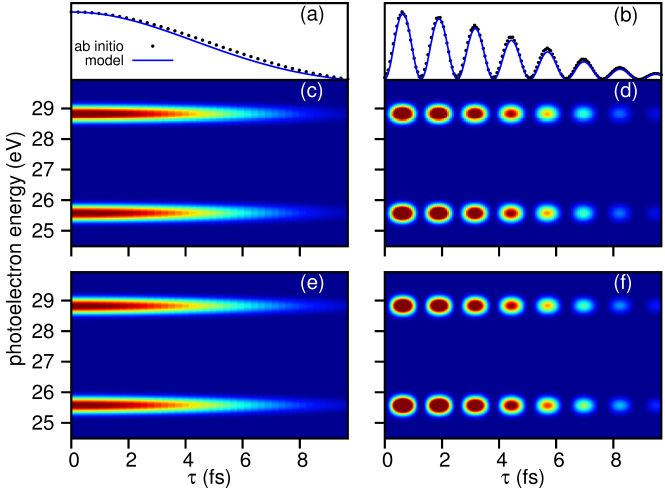}
\caption{\label{fig:PumpProbeSAPTraceHydrogen} (Color online) Spectrum for the two-photon ionization of the hydrogen atom from the ground state by means of a single (left panels) or a train of (right panels) XUV Gaussian pulses, in association with a 10~fs long IR probe pulse. Top panels (a,b): comparison between the energy integrated signal of the upper sideband, as a function the pump-probe time delay, computed \emph{ab initio} (black dotted line) or with the model (blue solid line). Middle panels (c,d): energy and time-delay resolved spectra from the \emph{ab initio} calculation (only states of even symmetry are shown). Bottom panels (e,f): energy and time-delay resolved spectra computed with the model.}
\end{center}
\end{figure}
The left panels of Figure~\ref{fig:PumpProbeSAPTraceHydrogen} illustrate the photoelectron spectrum of the hydrogen atom ionized from the ground state with a single XUV Gaussian pulse with duration of $5$~fs and central frequency 40.8~eV, in association with an 760~nm IR probe pulse 10~fs long, with an intensity of $10$GW/cm$^2$, as a function of the pump-probe time delay. The spectrum in the central panel is obtained \emph{ab initio} by solving the TDSE for the atom in a numerical basis, while the bottom panel is computed using Eq.~\eqref{eq:NonResonantFiniteAmplitude}. The spectrum computed with the model, which includes all the terms proportional to the intensity of the probe laser, accurately reproduces all the features in the real energy-integrated (Fig.~\ref{fig:PumpProbeSAPTraceHydrogen}a) and energy-resolved (Fig.~\ref{fig:PumpProbeSAPTraceHydrogen}c) spectrum. 

\paragraph{Red shift of RABITT beating with finite pulses.}\label{sec:IIe}

\begin{figure}[hbtp!]
\begin{center}
\includegraphics[width=0.9\linewidth]{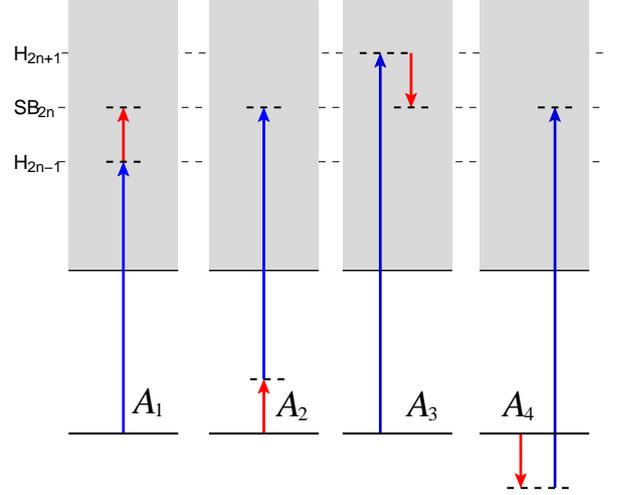}
\caption{\label{fig:RABITTscheme} (Color online) Quantum paths contributing to a sideband signal in RABITT spectroscopy. The amplitudes of both paths 1 and 2, in which one IR photon is absorbed, are modulated by a phase factor $e^{i\omega_\IR\tau}$, while those of paths 3 and 4 are modulated by a phase factor $e^{-i\omega_\IR\tau}$. As a result of the interference between the four amplitudes, therefore, the sideband signal beats with angular frequency $2\omega_\IR$.}
\end{center}
\end{figure}
In RABITT spectroscopy, the amplitude of each sideband SB$_{2n}$ is given by the sum of four time-ordered two-photon amplitudes,
\begin{equation}
\mathcal{A}_{\SB_{2n}}=\mathcal{A}_{\SB_{2n}}^{\HH_{2n-1}\IR}+\mathcal{A}_{\SB_{2n}}^{\IR\HH_{2n-1}}+\mathcal{A}_{\SB_{2n}}^{\HH_{2n+1}\bar{\IR}}+\mathcal{A}_{\SB_{2n}}^{\bar{\IR}\HH_{2n+1}},
\end{equation}
where $\bar{\IR}$ indicates the emission component (negative frequency) of the IR pulse (see Fig.~\ref{fig:RABITTscheme}). In the limit of long pulses, the frequency of the RABITT beating is $2\omega_\IR$ and the atomic phaseshift in the standard expression for the sideband intensity, $I_{\SB_{2n}}=I_0\,\cos(2\omega_\IR\tau+\Delta\phi_\HH+\Delta\varphi_\mathrm{at})$~\cite{Veniard1996}, is $\Delta\varphi_{at}=\arg[\mathcal{M}_{f i}(\omega_{2n-1})+\mathcal{M}_{f i}(\omega_\IR)]-\arg[\mathcal{M}_{f i}(\omega_{2n+1})+\mathcal{M}_{f i}(-\omega_\IR)]$. This is still the case if only one of either the APT or the IR has a long duration. Indeed, the RABITT frequency is given by the sum of the frequencies of the absorbed and of the emitted IR photons. If the XUV train comprises only multiples of the fundamental frequency $\omega_\IR$ or if the probe pulse is monochromatic with frequency $\omega_\IR$, then the only possible outcome for the RABITT beating is $2\omega_\IR$. 
\begin{figure}[hbtp!]
\begin{center}
\includegraphics[width=0.9\linewidth]{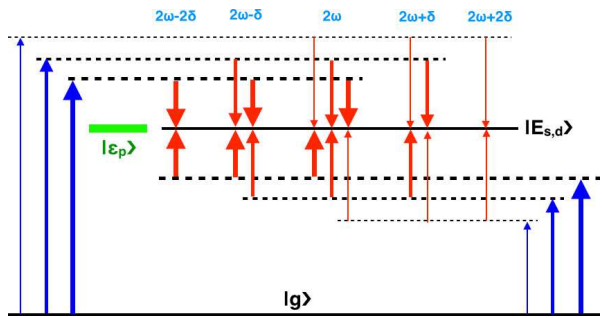}
\caption{\label{fig:TwoPhotonPolychrom} (Color online)
The XUV + IR above-threshold ionization amplitude is inversely proportional to the frequency of the IR photon. When both the XUV train and the IR pulse have finite duration, therefore, the signal is biased in favor of the low IR-frequency components. As a result, the spectrum of sideband beating in RABITT is red-shifted compared to the nominal $2\omega_\IR$ value.}
\end{center}
\end{figure}
On the other hand, when both the APT and the IR have finite duration, the RABITT beating is red-shifted with respect to the nominal $2\omega_\IR$ value. This is because non-resonant two-photon matrix elements, which are dominated by contributions from virtual states at the same energy as the final state, are inversely proportional to the energy of the last-exchanged IR photon, $\mathcal{M}_{E,i}\propto (E_i+\omega_{2n\pm1}-E_{2n})^{-1}=\pm\omega_\IR^{-1}$. Therefore, of the many IR wavelengths that contribute to the transition with finite pulses, long ones weigh more, thus biasing the RABITT beating towards the red (see Fig.~\ref{fig:TwoPhotonPolychrom}). The right central and bottom panels of Figure~\ref{fig:PumpProbeSAPTraceHydrogen} show the comparison between \emph{ab initio} and model calculations in the case of the RABITT ionization of the hydrogen atom, where a 5~fs long (fwhm) Gaussian  APT formed by Gaussian XUV pulses with central frequency of 40.8~eV and duration of 250~as is used in association with a 760~nm, 10~fs, 10GW/cm$^2$ probe pulse. 
For these pulse parameters, the Fourier transform of the sideband oscillation in Figure~\ref{fig:PumpProbeSAPTraceHydrogen}b, reveals a beating frequency which is red-shifted with respect to the nominal RABITT frequency by an amount of $0.021$~eV for the \emph{ab initio}, in good agreement with the value of $0.017$~eV predicted by the model. Part of the difference between these two values is explained by the use, in the \emph{ab initio} calculation, of a probe pulse with cosine-squared instead of Gaussian envelope, which permits us to reduce the size of the quantization box.

\section{Two-photon resonant model}\label{sec:ResonantModel}

In this section we will use Fano formalism to compute two-photon transition amplitudes for the case in which the intermediate and/or final continuum states feature isolated resonances. After a short overview of Fano's main results and a discussion of the phase properties of the one-photon Fano transition matrix element, which will be relevant for the following of this section, we will consider two-photon transition amplitudes for the case of single-channel continuum states with at most one isolated resonance. The generalisation to multiple single-channel isolated resonances will be straightforward. At the end of this section, we will discuss the extension of the model to the multichannel resonant case.

\subsection{Phase properties of Fano transition amplitude}
In stationary conditions, metastable states manifest themselves in single photoionization spectra as characteristically asymmetric peaks~\cite{Madden1963}. The asymmetry of experimental resonant profiles can be explained with the well-known Fano's formalism~\cite{Fano1961}. In the simplest formulation of Fano's approach, the field-free hamiltonian $H$ is given by the sum of an unperturbed component $H_0$ and a ``configuration interaction" component $V$, $H=H_0+V$, where the eigenstates of $H_0$ comprise a featureless continuum $|\varepsilon\rangle$ and a bound state $|a\rangle$, $H_0|\varepsilon\rangle=|\varepsilon\rangle\varepsilon$,  $H_0|a\rangle=|a\rangle E_a$, while the configuration interaction only couples the bound state to the continuum, $V_{a\epsilon}=\langle a | H-H_0|\epsilon\rangle$.  If the coupling $V_{a\epsilon}$ depends only weakly on the continuum index $\epsilon$, the continuum eigenstates of the full hamiltonian, $H|\psi_E\rangle=|\psi_E\rangle E$, can be expressed as
\begin{equation}\label{eq:Fano}
|\psi_{E}\rangle = |E\rangle+ \Bigg[
|a\rangle+\int d\varepsilon |\varepsilon\rangle \frac{V_{\varepsilon a}}{E-\varepsilon+i0^+}\Bigg]\frac{V_{aE}}{E-\tilde{E}_a},
\end{equation}
where $\tilde{E}_a(E)=E_a+ \Delta_a(E)-i\Gamma_a(E)/2$ is a complex function of the energy, with the so-called energy shift $\Delta_a(E)$ and width $\Gamma_a(E)$ defined as $\Delta_a(E)=P\sum\hspace{-10pt}\int d\varepsilon|V_{\varepsilon a}|^2/(E-\varepsilon)$ and $\Gamma_a(E)=2\pi^2|V_{Ea}|^2$. The pole of $E-\tilde{E}_a(E)$ in the negative complex plane is, by convention, the complex resonance energy. Notice that the solution~\eqref{eq:Fano}, which is readily obtained by projecting the Lippmann-Schwinger equation $|\psi_E\rangle = |E\rangle + G_0^+(E)V|\psi_E\rangle$ on the basis of unperturbed states, differs from Fano's original solution by a complex normalisation factor. Here we will use the form~\eqref{eq:Fano} because it is normalised, $\langle\psi_E|\psi_{E'}\rangle=\delta(E-E')$, and continuous with respect to $E$. The energy shift $\Delta_a(E)$ and width $\Gamma_a(E)$ depend only weakly on the energy $E$, so that one can assume they are constant in the energy region of interest, $\Gamma_a(E)\simeq\Gamma_a(E_a)=\Gamma_a$, $\Delta_a(E)\simeq\Delta_a(E_a)=\Delta_a$. In these conditions, which we assume to be fulfilled, the complex resonance energy is thus well approximated as $\tilde{E}_a\simeq E_a+\Delta_a-i\Gamma_a/2$. For our convenience, we will indicate the real part of the resonance energy as $\bar{E}_a=\mathrm{Re}[\tilde{E}_a]=E_a+\Delta_a$. It is customary to define a reduced energy variable $\epsilon=2(E-\bar{E}_a)/\Gamma_a$, and a distorted continuum component $|\tilde{a}\rangle\equiv|a\rangle+P\sum\hspace{-10pt}\int d\varepsilon|\varepsilon\rangle V_{\varepsilon, a}/(E-\varepsilon)$ which incorporates the original bound state $|a\rangle$. Using this notation, Eq.~\eqref{eq:Fano} can be reformulated as
\begin{equation}\label{eq:FanoSolution2}
|\psi_E\rangle = |E\rangle \frac{\epsilon}{\epsilon+i}+ |\tilde{a}\rangle\frac{1}{\pi V_{Ea}}\frac{1}{\epsilon+i}.
\end{equation}
Notice that for $\epsilon\to\pm\infty$, $|\psi_E\rangle$ converges to $|E\rangle$. We also define a resonant phaseshift $\phi_E$ as
\begin{equation}\label{eq:PhaseShift}
\phi_E\equiv \pi/2+\arctan\epsilon,
\end{equation}
which is a \emph{continuous, monotonically increasing} function of $E$, with $\phi_{-\infty}=0$, $\phi_{\infty}=\pi$.  If the $|E\rangle$ channel functions are used as reference asymptotes, $\phi_E$ is associated to the \emph{on-shell} scattering matrix $s(E)$ for the collisional excitation of the resonance,
\begin{equation}\label{eq:ScatteringMatrix}
s(E)=e^{2i\phi}=(\epsilon-i)/(\epsilon+i).
\end{equation}
Conversely, $\epsilon=-\cos\phi/\sin\phi$ (we will drop the energy suffix from $\phi_E$, when $E$ is clear from the context).  In this formalism, the dipole transition matrix element between an initial ground state $|g\rangle$ and a final resonant continuum $|\psi_E\rangle$ can be written as
\begin{equation}\label{eq:FanoTransitionAmplitude}
\langle\psi_E|\mathcal{O}|g\rangle=\mathcal{O}_{Eg}\frac{\epsilon+q_{\tilde{a}g}}{\epsilon-i},\qquadÊ\mathcal{O}_{Eg}=\langleÊE|\mathcal{O}|g\rangle,
\end{equation}
where $q_{\tilde{a}g}$ is a real parameter that measures the relative strength of the transition from the ground state to the autoionizing state, relative to that of the direct-ionization process,
\begin{equation}
q_{\tilde{a}g}=\frac{\mathcal{O}_{\tilde{a}g}}{\pi V_{aE}\mathcal{O}_{Eg}}.
\end{equation}

\begin{figure}[hbtp!]
\begin{center}
\includegraphics[width=\linewidth]{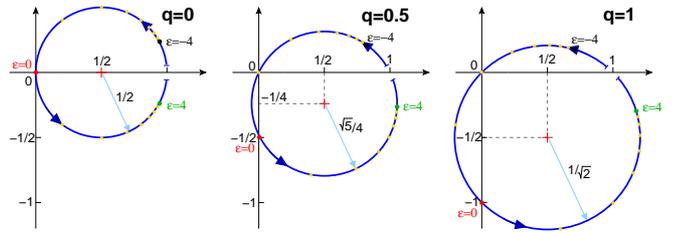}
\caption{(Color online) \label{fig:FanoAmplitude} Trajectory in the complex plane of the $\mathcal{R}(\epsilon)$ resonant factor in Fano's dipole transition amplitude, as the reduced detuning $\epsilon$ varies from large negative to large positive values. The three panels correspond to different values of the $q$ parameter. See text for more details.}
\end{center}
\end{figure}

The resonant factor $\mathcal{R}(\epsilon)=(\epsilon+q_{\tilde{a}g})/(\epsilon+i)$ in the complex conjugate of the dipole transition amplitude~\eqref{eq:FanoTransitionAmplitude} can be written as the sum of a constant term plus a second term proportional to the unimodular function $(\epsilon-i)/(\epsilon+i)$,
\begin{equation}\label{eq:CircularFanoAmplitude1}
\mathcal{R}(\epsilon)\,=\,\frac{\epsilon+q}{\epsilon+i}\,=\,\frac{1-iq}{2}\,+\,\frac{1+iq}{2}\,\frac{\epsilon-i}{\epsilon+i}\,.
\end{equation}
This means that, as $\epsilon$ increases from $-\infty$ to $+\infty$, $\mathcal{R}(\epsilon)$ describes counterclockwise a circle in the complex plane. The circle is centered at $(1-iq)/2$, it has radius $r=\sqrt{1+q^2}\,/\,2$, and it both starts and ends at $1$, intercepting the origin at $\epsilon=-q$.  Figure~\ref{fig:FanoAmplitude} illustrates the trajectory of $R(\epsilon)$, from large negative ($\epsilon \ll -1)$ to large positive detunings ($\epsilon\gg1$), for three representative values of $q$: $0$, $0.5$, and $1$.
This geometrical interpretation can be visualized even more clearly by defining the angular variable $\varphi=\arctan(q)\in(-\pi/2,\pi/2)$, equivalent to the one introduced by Ott~\emph{et al.}~\cite{Ott2013} in the context of the dipolar response of a Fano resonance, with which $\mathcal{R}(\epsilon)$ becomes
\begin{equation}\label{eq:CircularFanoAmplitude1}
\mathcal{R}(\epsilon)\,=\,r\,e^{-i\varphi}\,+\,r\,e^{i\varphi}\,e^{2i\phi}.
\end{equation}
Notice that even if the resonant dipole transition amplitude~\eqref{eq:FanoTransitionAmplitude} is a continuous function of $\epsilon$, its phase is not. Indeed, the latter experiences a discontinuous jump of $\pi$ in correspondence of $\epsilon=-q$, when the circle intercepts the origin,
\begin{eqnarray}
\argÊ\mathcal{R(\epsilon)}
 & = & \arg\left\{e^{i\phi}\left[e^{i(\phi+\varphi)}+e^{-i(\phi+\varphi)}\right]\right\}=\\
 & = & \phi + \arg\left[\cos(\phi+\varphi)\right] =\\
 & = & \arctan(\epsilon) - \pi \theta(\epsilon+q).
\end{eqnarray}
Far from reflecting a real discontinuity in the physical properties of the system, this circumstance simply reflects the fact that $\arg(z)$ is discontinuous at the origin.

The photoionization cross section $\sigma(E)$, which is proportional to the square module of the dipole transition amplitude to the resonant continuum, therefore, is the product between a background smooth cross section $\sigma_{\mathrm{bg}}(E)$, associated to direct photoionization, and the celebrated Fano profile,
\begin{equation}\label{eq:FanoProfileInSecIII}
\sigma(E)=\sigma_{\mathrm{bg}}(E)\frac{(\epsilon+q_{\tilde{a}g})^2}{\epsilon^2+1}.
\end{equation}
In this latter expression, any information on the relative phase between different frequency components of the photoelectron wavepacket generated by the interaction of the impinging ionizing light is lost. Therefore, while Equation~\eqref{eq:FanoProfileInSecIII} is sufficient to interpret one-photon ionization experiments such as those conducted at synchrotron facilities, when two or more photons are exchanged in a coherent transition, the relative phase of intermediate transition amplitudes becomes essential and we must go back to Eq.~\eqref{eq:FanoTransitionAmplitude} instead.

\subsection{Resonant two-photon transition matrix element.}\label{sec:IIIB}
To derive the analytical formula for finite-pulse resonant two-photon transition amplitudes, we first need to obtain an approximated analytical expression for the two-photon ionization matrix element $\mathcal{M}_{\betaÊE,g}(\omega)$, 
\begin{equation}\label{eq:TransMatElem}
\mathcal{M}_{\beta E,g} (\omega) = \sumint d\varepsilon \frac{\langle\psi_{\beta E} | \mathcal{O} | \psi_{\alpha \varepsilon}\rangle\,\langle\psi_{\alpha\varepsilon}| \mathcal{O} | g\rangle}{E_g+\omega-\varepsilon + i0^+}.
\end{equation}
To do so, we assume that the continuum branches in both the intermediate states, $|\psi_{\alpha E}\rangle$, and final states, $|\psi_{\beta E}\rangle$, can be expressed, using Fano's formalism, in terms of known bound and continuum eigenstates of a reference hamiltonian $H_0$, $H_0|a\rangle=E_a|a\rangle$, $H_0|b\rangle=E_b|b\rangle$, $H_0|\gamma \varepsilon\rangle=\varepsilon|\gamma \varepsilon\rangle$, 
\begin{eqnarray}
|\psi_{\alpha E}\rangle &=& |\alpha E\rangle + 
\left(
|a\rangle+\hspace{-2pt}\int\hspace{-2pt} \frac{d\varepsilon|\alpha\varepsilon\rangle\,V_{\alpha\varepsilon, a}}{E-\varepsilon+i0^+}
\right) \frac{V_{a,\alpha E}}{E-\tilde{E}_a},\label{eq:FanoIntermediate}\\
|\psi_{\beta E}\rangle &=& |\beta E\rangle + 
\left(
|b\rangle+\hspace{-2pt}\int\hspace{-2pt} \frac{d\varepsilon|\beta\varepsilon\rangle\,V_{\beta\varepsilon, b}}{E-\varepsilon+i0^+}
\right) \frac{V_{b,\beta E}}{E-\tilde{E}_b},\label{eq:FanoFinal}
\end{eqnarray}
where $V$ denotes the field-free electron-electron interaction not included in $H_0$, e.g., $V_{a,\alpha \varepsilon}=\langle a|H-H_0|\alpha\varepsilon\rangle$. The interacting-continuum wavefunctions in Eqns.~\eqref{eq:FanoIntermediate} and~\eqref{eq:FanoFinal} are normalized as $\langle\psi_{E'}| \psi_{E}\rangle = \delta(E'-E)$.
The suffixes $\alpha$ and $\beta$ identify the ionization channel in the intermediate and final states, respectively, i.e., the collection of discrete quantum numbers needed to specify the asymptotic state of the parent ion, of the photoelectron, as well as their angular and spin coupling.  Let us separate in $\mathcal{M}_{\beta E,g}(\omega)$ the contribution of the intermediate bound states $\{|n\rangle\}$, $\mathcal{M}^{(b)}_{\beta E,g}(\omega)$, from that of the intermediate continuum states $|\psi_{\alpha E}\rangle$, $\mathcal{M}^{(c)}_{\beta E,g}(\omega)$, 
\begin{eqnarray}
\mathcal{M}_{\beta E,g} (\omega) &=& \mathcal{M}^{(b)}_{\beta E, g}(\omega)+\mathcal{M}^{(c)}_{\beta E,g}(\omega),\label{eq:splitTransMatElem}\\
\mathcal{M}^{(b)}_{\beta E,g}(\omega)&=&\sum_n
\frac{\langle\psi_{\beta E} | \mathcal{O} |n\rangle\, \mathcal{O}_{ng} }{\omega-\omega_{ng}+i0^+}\\
\mathcal{M}^{(c)}_{\beta E,g}(\omega)&=&\int d\varepsilon \frac{\langle\psi_{\beta E} | \mathcal{O} | \psi_{\alpha \varepsilon}\rangle\,\langle\psi_{\alpha\varepsilon}| \mathcal{O} | g\rangle}{E_g+\omega-\varepsilon + i0^+}.\label{eq:Mc}
\end{eqnarray}
The transition matrix elements between a localised state and a Fano continuum can be accurately parametrised with Fano's formula, 
\begin{eqnarray}
\langle\psi_{\alpha \varepsilon}|\mathcal{O}|g\rangle &=& \frac{\epsilon_{\varepsilon a}+q_{\tilde{a}g}}{\epsilon_{\varepsilon a}-i}\mathcal{O}_{\alpha \varepsilon, g},
\label{eq:FanoAmp_ag}\\
\langle\psi_{\beta E}|\mathcal{O}|n\rangle &=& \frac{\epsilon_{Eb}+q_{\tilde{b}n}}{\epsilon_{Eb}-i}\mathcal{O}_{\beta E, n}.
\end{eqnarray}
The contribution from bound intermediate states, therefore, is readily written as
\begin{equation}
\mathcal{M}^{(b)}_{\beta E,g}(\omega)=\sum_n \frac{\epsilon_{Eb}+q_{\tilde{b}n}}{\epsilon_{Eb}-i}
\frac{\mathcal{O}_{\beta E, n}\mathcal{O}_{ng} }{\omega-\omega_{ng}+ i0^+}.\label{eq:Mb1}
\end{equation}
In practical cases, this expression can often be restricted to the contribution from a limited set of intermediate bound states, or even from just one of them. For example, in the excitation of helium from the $1s^2$ ground state to the doubly excited states with $N=2$, the biggest role in Eq.~\eqref{eq:Mb1} is played by the intermediate $1s2p$ state, for which the oscillator strength with the $N=2$ states is very large and the background ionization amplitude is very small ($q_{\tilde{b}n}\gg1$. The 2p$^2 \gets$ 1s2p $\gets$ 1s$^2$ is a characteristic example). In this case, if the intermediate bound state is non-resonant, one can use the simplified expression
 \begin{equation}
\mathcal{M}^{(b)}_{\beta E,g}(\omega)\approx \frac{q_{\tilde{b}n_0}}{\epsilon_{Eb}-i}
\frac{\mathcal{O}_{\beta E, n_0}\mathcal{O}_{n_0,g} }{\omega-\omega_{n_0g}}.
\end{equation}
The latter expression is applicable even in the case of multiple intermediate states that contribute to the transition amplitude by means of virtual excitations and which are clustered in an energy region that is small if compared with the detuning $\omega-\omega_{n_0g}$ from the absorption of the first photon. Let us now consider the contribution from the intermediate continuum states.  Replacing eq.~\eqref{eq:FanoAmp_ag} in~\eqref{eq:Mc} we find
\begin{equation}
\mathcal{M}^{(c)}_{\beta E,g}(\omega)=\int  \frac{d\varepsilon\langle\psi_{\beta E} | \mathcal{O} | \psi_{\alpha \varepsilon}\rangle}{E_g+\omega-\varepsilon + i0^+}
\frac{\epsilon_{\varepsilon a}+q_{\tilde{a}g}}{\epsilon_{\varepsilon a}-i}\mathcal{O}_{\alpha \varepsilon, g}.
\label{eq:Mc2}
\end{equation}
To advance further, we must find an expression for the continuum-continuum resonant transition amplitude $\langle\psi_{\beta E} | \mathcal{O} | \psi_{\alpha \varepsilon}\rangle$ in terms of a limited number of almost-constant parameters. In analogy with the Fano formula for the dipole transition from bound states, we first take out from this matrix element the term that involves only transition matrix elements between states in the unperturbed continuum,
\begin{eqnarray}
&&\langle\psi_{\beta E} | \mathcal{O} | \psi_{\alpha \varepsilon}\rangle=\langle\bar{\psi}_{\beta E}|\mathcal{O}|\bar{\psi}_{\alpha \varepsilon}\rangle-\frac{V_{\beta E,b}}{E-\tilde{E}_b^*}\mathcal{O}_{ba}\frac{V_{a,\alpha \varepsilon}}{\varepsilon-\tilde{E}_a}+\nonumber\\
&&+\frac{V_{\beta E,b}}{E-\tilde{E}_b^*}\langle b|\mathcal{O}|\psi_{\alpha\varepsilon}\rangle + \langle\psi_{\beta E}|\mathcal{O}|a\rangle\frac{V_{a,\alpha \varepsilon}}{\varepsilon-\tilde{E}_a}
=\nonumber\\
&=&\langle\bar{\psi}_{\beta E}|\mathcal{O}|\bar{\psi}_{\alpha \varepsilon}\rangle+\frac{1}{\pi}
\frac{1}{\epsilon_{Eb}-i}\frac{1}{\epsilon_{\varepsilon a}+i}
\Big[
-\frac{\mathcal{O}_{ba}}{\pi V_{b,\beta E}V_{\alpha \varepsilon,a}}+\nonumber\\
&+&\frac{\mathcal{O}_{b,\alpha\varepsilon}}{V_{b,\beta E}}(\epsilon_{\varepsilon a}+q_{\tilde{a}b})+
\frac{\mathcal{O}_{\beta E, a}}{V_{\alpha \varepsilon,a}}(\epsilon_{Eb}+q_{\tilde{b}a})\Big],\label{eq:FullCCTME}
\end{eqnarray} 
where the barred states represent the Fano continuum without the bound component, i.e., 
\begin{equation}\label{eq:}
|\bar{\psi}_{\beta E}\rangle= |\beta E\rangle + \int \frac{d\varepsilon|\beta\varepsilon\rangle\,V_{\beta\varepsilon, b}}{E-\varepsilon+i0^+}\frac{V_{b,\beta E}}{E-\tilde{E}_b},
\end{equation} 
and we used the relation $\Gamma_a=2\pi |V_{a,\alpha E}|^2$. By applying the \emph{on-shell} approximation, and assuming that $\bar{\mathcal{O}}_{\alpha\beta}\equiv\bar{\mathcal{O}}_{\alpha\beta}(E)$, $V_{a,\alpha E}$, and $V_{b,\beta E}$ are sufficiently slowly varying functions of $E$, it is easy to show that 
\begin{eqnarray}
\langle\bar{\psi}_{\beta E}|\mathcal{O}|\bar{\psi}_{\alpha \varepsilon}\rangle&=&
\bar{\mathcal{O}}_{\beta\alpha}\delta(E-\varepsilon)+\label{eq:BarredTME}\\
&+&\frac{1}{\pi}\frac{\bar{\mathcal{O}}_{\beta\alpha}}{\varepsilon-E+i0^+}\frac{\epsilon_{Eb}-\epsilon_{\varepsilon a}}{(\epsilon_{\varepsilon a}+i)(\epsilon_{E b}-i)}.\nonumber
\end{eqnarray}
Indeed, to compute the transition matrix element between the two modified continua, it is sufficient to close the integration path with a semi-circular path in either the upper or the lower half complex plane, where the argument of the integral decreases quadratically with respect to the integration variable, and apply Cauchy residual theorem. By combining Eqs.~\eqref{eq:FullCCTME} and \eqref{eq:BarredTME}, the dipole transition amplitudes between the two Fano resonant continua can be approximated as
\begin{equation}
\begin{split}
&\langle\psi_{\beta E} | \mathcal{O} | \psi_{\alpha \varepsilon}\rangle=\bar{\mathcal{O}}_{\beta\alpha}\delta(E-\varepsilon)+\\
&+
\frac{\bar{\mathcal{O}}_{\beta\alpha}}{\varepsilon-E+i0^+}\frac{\epsilon_{Eb}-\epsilon_{\varepsilon a}}{\pi(\epsilon_{\varepsilon a}+i)(\epsilon_{E b}-i)}+
\\
&+
\frac{
\frac{\mathcal{O}_{b,\alpha\varepsilon}(\epsilon_{\varepsilon a}+q_{\tilde{a}b})}{V_{b,\beta E}}+ 
\frac{\mathcal{O}_{\beta E,a}(\epsilon_{Eb}+q_{\tilde{b}a})}{V_{\alpha \varepsilon,a}} -
\frac{\mathcal{O}_{ba}}{\pi V_{\alpha \varepsilon,a}V_{b,\beta E}}}
{\pi(\epsilon_{\varepsilon a}+i)(\epsilon_{E b}-i)}.
\end{split}
\end{equation} 
We can now insert this expression in the continuum contribution~\eqref{eq:Mc2} to the two-photon matrix element. The integral of the argument proportional to a Dirac delta function is evaluated immediately, while the other two terms require a more careful discussion,
\begin{equation}
\mathcal{M}^{(c)}_{\beta E,g}(\omega)=
\frac{\bar{\mathcal{O}}_{\beta\alpha}\mathcal{O}_{\alpha E, g}}{E_g+\omega-E + i0^+}
\frac{\epsilon_{E a}+q_{\tilde{a}g}}{\epsilon_{E a}-i}+I_2+I_3,
\end{equation}
where
\begin{equation}
I_2=\frac{1/\pi}{\epsilon_{E b}-i}\int
\frac{\epsilon_{\varepsilon a}+q_{\tilde{a}g}}{\epsilon_{\varepsilon a}^2+1}
\frac{\epsilon_{\varepsilon a}-\epsilon_{Eb}}{\varepsilon-E+i0^+}\frac{\bar{\mathcal{O}}_{\beta\alpha}\mathcal{O}_{\alpha \varepsilon, g}d\varepsilon}{\varepsilon-E_g-\omega-i0^+}\nonumber
\end{equation}
and
\begin{eqnarray}
I_3&=&\int \frac{\mathcal{O}_{\alpha \varepsilon, g}d\varepsilon}{E_g+\omega-\varepsilon + i0^+}
\frac{\epsilon_{\varepsilon a}+q_{\tilde{a}g}}{\epsilon_{\varepsilon a}-i}\times\label{eq:DefI3}\\
&\times&
\frac{
\frac{\mathcal{O}_{b,\alpha\varepsilon}(\epsilon_{\varepsilon a}+q_{\tilde{a}b})}{V_{b,\beta E}}+ 
\frac{\mathcal{O}_{\beta E,a}(\epsilon_{Eb}+q_{\tilde{b}a})}{V_{\alpha \varepsilon,a}}-
\frac{\mathcal{O}_{ba}}{\pi V_{b,\beta E}V_{\alpha \varepsilon,a}}
}{\pi (\epsilon_{\varepsilon a}+i)(\epsilon_{E b}-i)}.\nonumber
\end{eqnarray}
For large values of $\varepsilon$, the argument of the integral in $I_2$ is inversely proportional to $\varepsilon^{2}$. Therefore, this integral can be conveniently computed by closing the integration circuit with a semi-circular arc in the lower half of the complex plane, provided that the transition matrix elements are only weakly varying on the additional arc, for large enough arc radii. The result is
\begin{eqnarray}
I_2&=&\frac{\epsilon_{E a}-\epsilon_{E b}}{(\epsilon_{E a}-i)(\epsilon_{E b}-i)}
\frac{\epsilon_{E a}+q_{\tilde{a}g}}{\epsilon_{E a}+i}
\frac{2i\bar{\mathcal{O}}_{\beta\alpha}\mathcal{O}_{\alpha E, g}}{\omega-E+E_g+i0^+}
-\nonumber\\
&&-\frac{\epsilon_{Eb}+i}{\epsilon_{E b}-i}\frac{q_{\tilde{a}g}-i}{\epsilon_{E a}+i}
\frac{\bar{\mathcal{O}}_{\beta\alpha}\mathcal{O}_{\alpha E, g}}{\omega-\omega_{\tilde{a}g}}.
\end{eqnarray}

The last integral, $I_3$, has only one simple pole in the lower complex plane and hence it also would be conveniently computed by closing the integration circuit in the lower half of the complex plane with the arc $\Gamma_R=\{R e^{-i\varphi},\varphi\in[0,\pi]\}$,
\begin{equation}
I_3=\lim_{R\to\infty}\left[\int_{[-R,R]\cup\Gamma_R}\hspace{-35pt}\mathcal{I}_3(z)dz-\int_{\Gamma_R}\hspace{-5pt}\mathcal{I}_3(z)dz\right],
\end{equation}
where $\mathcal{I}_3(\varepsilon)$ indicates the argument of the integral in~\eqref{eq:DefI3}. In contrast to the previous case, however, the absolute value of $\mathcal{I}_3(\varepsilon)$ decays only as $|\varepsilon|^{-1}$. Instead of vanishing as $R\to\infty$, therefore, the contribution of the arc converges to a finite value that must be taken into account, and which is easily computed (as usual, we assume that all the matrix elements are constant in a region of the complex plane sufficiently large to attain reasonable convergence of the circuit integral), 
\begin{equation}
\lim_{R\to\infty}\int_{\Gamma_R}\hspace{-5pt}\mathcal{I}_3(z)dz=\frac{i\mathcal{O}_{b,\alpha E}\mathcal{O}_{\alpha E, g}}{V_{b,\beta E}(\epsilon_{E b}-i)}.
\end{equation}
The value of the total integral $I_3$, then, becomes
\begin{eqnarray}
I_3&=&-i\frac{\mathcal{O}_{b,\alpha}}{V_{b,\beta}}\frac{\mathcal{O}_{\alpha, g}}{\epsilon_{E b}-i}+
\frac{\Gamma_a}{2} \frac{q_{\tilde{a}g}-i}{\omega-\omega_{\tilde{a}g}}\frac{\mathcal{O}_{\alpha, g}}{\epsilon_{E b}-i}\times\\
&\times&\left[
\frac{\mathcal{O}_{b,\alpha}(q_{\tilde{a}b}-i)}{V_{b,\beta}}
+\frac{\mathcal{O}_{\beta,a}(\epsilon_{Eb}+q_{\tilde{b}a})}{V_{\alpha,a}}
-\frac{\mathcal{O}_{ba}/\pi}{V_{b,\beta}V_{\alpha,a}}
\right].\nonumber
\end{eqnarray}

In summary, the expression for the intermediate-continuum contribution to the two-photon resonant transition matrix element is
\begin{eqnarray}\label{eq:MComplete}
\mathcal{M}^{(c)}_{\beta E,g}(\omega)&=&
\frac{\epsilon_{E a}+q_{\tilde{a}g}}{\epsilon_{E a}+i}\frac{\epsilon_{E b}+i}{\epsilon_{E b}-i}
\frac{\bar{\mathcal{O}}_{\beta\alpha}\mathcal{O}_{\alpha, g}}{E_g+\omega-E + i0^+}-\nonumber\\
&-&\frac{q_{\tilde{a}g}-i}{\epsilon_{E a}+i}
\frac{\epsilon_{E b}+i}{\epsilon_{E b}-i}\frac{\bar{\mathcal{O}}_{\beta\alpha}\mathcal{O}_{\alpha, g}}{\omega-\omega_{\tilde{a}g}}+\nonumber\\
&+&
(q_{\tilde{a}b}-i)
\frac{q_{\tilde{a}g}-i}{\epsilon_{E b}-i}\frac{\Gamma_a/2}{V_{b,\beta E}}\frac{\mathcal{O}_{b,\alpha}\mathcal{O}_{\alpha, g}}{\omega-\omega_{\tilde{a}g}}+\\
&+&
\pi V_{a,\alpha}(\epsilon_{Eb}+q_{\tilde{b}a})
\frac{q_{\tilde{a}g}-i}{\epsilon_{E b}-i}\frac{\mathcal{O}_{\beta,a}\mathcal{O}_{\alpha, g}}{\omega-\omega_{\tilde{a}g}}-\nonumber\\
&-&
\frac{q_{\tilde{a}g}-i}{\epsilon_{E b}-i}\frac{V_{a,\alpha}}{V_{b,\beta}}\frac{\mathcal{O}_{ba}\mathcal{O}_{\alpha, g}}{\omega-\omega_{\tilde{a}g}}
-\frac{i\mathcal{O}_{b,\alpha}\mathcal{O}_{\alpha, g}}{V_{b,\beta}(\epsilon_{E b}-i)}.\nonumber
\end{eqnarray}
This approximate algebraic expression for the two-photon transition matrix element in the presence of both an intermediate and a final autoionizing states is one of the main results of the present work. 

\begin{figure}[hbtp!]
\begin{center}
\includegraphics[width=\linewidth]{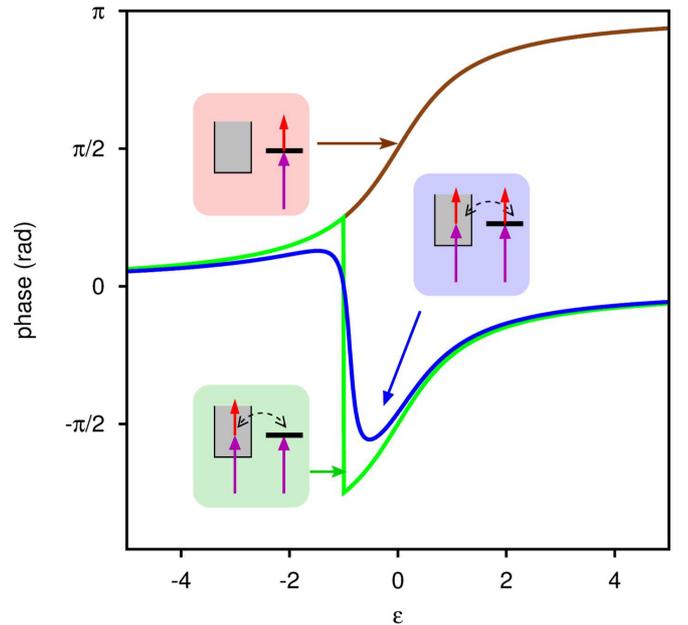}
\caption{\label{fig:FanoPhase} (Color online). Argument of the resonant factor $\mathcal{R}^{21}$ \eqref{eq:TPResonantFactor} of the two-photon matrix element~\eqref{eq:simpleMIntRes1} as a function of the reduced detuning of the pump photon from the intermediate resonance. Brown line: the ground state is radiatively coupled to the discrete but not to the homogeneous component of the intermediate state, i.e., $q \rightarrow \infty$. Green line: the ground state is radiatively coupled to both the discrete and homogeneous components ($q=1$), but the discrete component is not radiatively coupled to the final state, i.e., $\beta_{Ea} \rightarrow 0$. Blue line: both the discrete and continuum intermediate components are radiatively coupled with the initial and final states.}
\end{center}
\end{figure}

\paragraph{Case of no final resonances.}
In the relevant special case in which there are no final resonances, Eq.~\eqref{eq:MComplete} simplifies considerably since one can take its limit for  vanishing radiative and non-radiative couplings involving  the $|b\rangle$ state. The result is
\begin{eqnarray}
\mathcal{M}^{(c)}_{\beta E,g}(\omega)&=&
\frac{\epsilon_{E a}+q_{\tilde{a}g}}{\epsilon_{E a}+i}
\frac{\bar{\mathcal{O}}_{\beta\alpha}\mathcal{O}_{\alpha, g}}{E_g+\omega-E + i0^+}+\label{eq:MIntermediateResonance}\\
&+&\left(\beta_a-\frac{1}{\epsilon_{E a}+i}\right)
(q_{\tilde{a}g}-i)\frac{\bar{\mathcal{O}}_{\beta\alpha}\mathcal{O}_{\alpha, g}}{\omega-\omega_{\tilde{a}g}}\nonumber
\end{eqnarray}
where the parameter $\beta_a=\pi \mathcal{O}_{\beta,a}V_{a\alpha}/\bar{\mathcal{O}}_{\beta\alpha}$ is a pure number that depends solely on the properties of the atomic system. When considering resonant two-photon transitions with long overlapping pulses with frequencies $\omega_1$ and $\omega_2$ and duration larger than the lifetime of the intermediate resonance, Eq.~\eqref{eq:MIntermediateResonance} can be simplified further, since the conservation of energies applies, $E= E_g+\omega_1+\omega_2$. With few algebraic passages, it is easy to show that the matrix element appropriate for the time-ordered diagram in which photon $\omega_{1}$ is absorbed first becomes
\begin{equation}\label{eq:simpleMIntRes1}
\mathcal{M}^{(c,21)}_{\beta E,g}(\omega_1)=-\frac{\bar{\mathcal{O}}_{\beta\alpha}\mathcal{O}_{\alpha, g}}{\omega_2}\,\frac{\epsilon_{E_1a}+q_{\tilde{a}g}(1-\gamma_{a2})+i\gamma_{a2}}{\epsilon_{E_1a}+i},
\end{equation}
where we introduced the new real parameter
\begin{equation}
\gamma_{a2}=\frac{\omega_2\beta_a}{\Gamma_a/2}=\frac{\mathcal{O}_{\beta,a}}{\bar{\mathcal{O}}_{\beta\alpha}\frac{1}{\omega_2}V_{\alpha a}}
\end{equation}
which measures the relative strength of two alternative paths for the dipole transition from the intermediate bound state $|a\rangle$ to the final continuum $|\beta E\rangle$: a direct one, $\mathcal{O}_{\beta,a}$, and an indirect one, $\bar{\mathcal{O}}_{\beta\alpha}V_{\alpha a}/\omega_2$ in which the transition is mediated by the non-radiative coupling of the bound state with the intermediate continuum $|\alpha E\rangle$ followed by the dipole coupling between the intermediate and final continuum. Notice that in the formulation~\eqref{eq:simpleMIntRes1}, the reduced energy term $\epsilon$ is always relative to the energy of the intermediate state reached from the ground state by the absorption of the first photon, exactly as in the one-photon formula~\eqref{eq:FanoTransitionAmplitude}. It is interesting, therefore, to analyze more in detail the similarities and differences between expression~\eqref{eq:simpleMIntRes1} and that for one-photon transitions. First of all, if we define an effective $q$ parameter  as
\begin{equation}
q_{\mathrm{eff}}^{(21)}=q_{\tilde{a}g}(1-\gamma_{a2})+i\gamma_{a2},
\end{equation}
the resonant factor is formally the same in either expressions, 
\begin{equation}
\mathcal{R}^{(21)}=\frac{\epsilon_{E_1a} + q_{\mathrm{eff}}^{(21)}}{\epsilon_{E_1a}+i}.\label{eq:TPResonantFactor}
\end{equation}
Only in the case in which the intermediate bound state $|a	\rangle$ is not radiatively coupled to the final continuum ($\mathcal{O}_{\beta a}=0$ $\implies$ $q_{\mathrm{eff}}^{(21)}=q_{\tilde{a}g}$), however, do the resonant factors in the one-photon and the two-photon transition amplitude actually coincide in value and not in form only (see green line in Fig.~\ref{fig:FanoPhase}). In general, if $\gamma_{a2}\neq 0$, $q_{\mathrm{eff}}^{(21)}$ is a complex number which depends on the frequency of the second exchanged photon. The resonant factor in the two-photon transition matrix element can also be written as
\begin{equation}
\mathcal{R}^{(21)}=\gamma_{a2} + (1-\gamma_{a2})\frac{\epsilon_{E_1a} +q_{\tilde{a}g}}{\epsilon_{E_1a}+i},
\end{equation}
which is the same factor as in the one-photon case, scaled by $(1-\gamma_{a2})$ and shifted along the real axis by $\gamma_{a2}$. In particular, as the reduced detuning $\epsilon_{E_1a}$ is increased from $-\infty$ to $+\infty$, $\mathcal{R}^{(21)}(\epsilon_{E_1a})$ still describes counterclockwise a circle that starts and ends at $1$. In contrast to the one-photon case, however, if $\gamma_{a2}\neq 0$, the circle does not intersect the origin. In particular, if $\gamma_{a2}<0$, the circle, which is expanded with respect to the one-photon case, intersects the real axis at $\gamma_{a2}$ and at $1$, thus encircling the origin. This means that the phase of the two-photon transition matrix elements experiences a full $2\pi$ excursion. If, on the other hand, $\gamma_{a2}>0$, the circle is contracted and it misses the origin. In this latter case, the phase of the two-photon transition matrix elements experiences a finite excursion but no overall variation (see blue line in Fig.~\ref{fig:FanoPhase}).
Furthermore, since $\gamma_{a2}$ is proportional to $\omega_2$, the sign of $\gamma_{a2}$ for the emission of the second photon is the opposite of that for its absorption, the full $2\pi$ phase excursion and the no-net phase excursion case are both simultaneously present, one for the upper and one for the lower sidebands of the resonant two-photon transition. In the particular case in which $\gamma_{a2}=1$, the circular complex trajectory contracts to a point, $\mathcal{R}^{(21)}=1$, so that the two-photon amplitude does not bear any sign of the intermediate resonance (the amplitude for the opposite sideband, however, would still exhibit a pronounced $2\pi$ phase excursion).

So far, we have considered only the case in which the photon close to the resonance is absorbed first. In fact, the same final state is also reached by the path in which the photon $\omega_2$ is exchanged first, and whose matrix element is $\mathcal{M}_{\beta E,g}^{(c,12)}(\omega_2)$. In XUV-pump IR-probe experiment, where $\omega_{\IR}\ll\omega_{\XUV}$, and where the first excitation energy of the ground state is typically much larger than $\omega_\IR$, the contribution of the second path is generally small and it is often disregarded. Yet, the total transition matrix element should be computed as the sum of the two time-ordered contribution. If the path in which $\omega_2$ is exchanged first is not resonant, then we can imagine that this term contributes with a small complex constant to the total transition. In principle, therefore, the inverted-order transition has an effect similar to that of $\gamma_{a2}$, as it shifts the transition amplitude trajectory towards or away from the origin.

The limit in which only the intermediate state $|a\rangle$ is radiatively coupled to the ground, while the intermediate continuum is not ($q_{\tilde{a}g}=\infty$), is also interesting, since it effectively reproduces the assumptions that have been made in~\cite{Swoboda2010} and~\cite{Caillat2011}, and which lead to a neat $\pi$ excursion of the transition amplitude phase, as shown by the brown line in Fig.~\ref{fig:FanoPhase}. In the general case, where the discrete-continuum dipole coupling is not negligible, the typical abrupt $\pi$ discontinuity of~\eqref{eq:CircularFanoAmplitude1} disappears (see blue line in Fig.~\ref{fig:FanoPhase}). 

\paragraph{Case of no intermediate resonances.}
Two-photon excitation of a metastable state in the final continuum, with no intermediate resonances, which has been explored in the past by Cormier \emph{et al.}~\cite{Cormier1993}, is a second relevant case. 
The frequency-dependent two-photon matrix element for this case is readily obtained from the general formula~\eqref{eq:MComplete} by suppressing all the terms that involve the intermediate state $|a\rangle$,
\begin{eqnarray}\label{eq:MFinalResonance}
\mathcal{M}^{(c)}_{\beta E,g}(\omega)&=&
\frac{\epsilon_{E b}+i}{\epsilon_{E b}-i}
\frac{\bar{\mathcal{O}}_{\beta\alpha}\mathcal{O}_{\alpha, g}}{ E_g+\omega-E + i0^+}-
\frac{i\mathcal{O}_{b,\alpha}\mathcal{O}_{\alpha, g}}{V_{b,\beta}(\epsilon_{E b}-i)}.\nonumber
\end{eqnarray}
If we specialise this formula to the long-pulse limit, and assume the conservation of energy $E= E_g+\omega_1+\omega_2$, we obtain
\begin{eqnarray}\label{eq:MFinalResonance2}
\mathcal{M}^{(c,21)}_{\beta E,g}(\omega_1)&=&
-\frac{\bar{\mathcal{O}}_{\beta\alpha}\mathcal{O}_{\alpha, g}}{\omega_2}
\frac{\epsilon_{E b}+i(1+\gamma_{b2})}{\epsilon_{E b}-i},
\end{eqnarray}
where we introduced the new real parameter
\begin{equation}
\gamma_{b2}=\frac{\omega_2\mathcal{O}_{b,\alpha}}{V_{b,\beta}\bar{\mathcal{O}}_{\beta\alpha}}.
\end{equation}
A first surprising aspect of the resonant transition matrix element \eqref{eq:MFinalResonance2} in the present model is that it has a purely imaginary $q$ parameter, $q=i(1+\gamma_{b2})$. As mentioned at the beginning of this section, however, when autoionizing final states are involved, the contribution of intermediate virtual bound states can be very large and, when added to \eqref{eq:MFinalResonance2}, they give rise to an effective complex $q$ parameter with comparable real and imaginary components, as predicted in~\cite{Cormier1993}. Notice that, if the radiative coupling between the intermediate continuum and the final bound state is sufficiently large, it is in principle possible to select a value of $\omega_2$ such that $1+\gamma_{b2}$ vanishes, thus making the transition amplitude disappear at one of the final resonances, as it happens at $\epsilon=-q$ for a traditional Fano profile.

\subsection{Time-resolved transition amplitudes}\label{sec:IIIC}
From the expressions for the continuum~\eqref{eq:MComplete} and discrete~\eqref{eq:Mb1} contribution to the resonant two-photon transition matrix element, we can now proceed to compute the full transition amplitude associated to a pair of Gaussian pump and probe pulses. To do so, we will fold the transition matrix element $\mathcal{M}(\omega)$ with the FT of the field, as prescribed in Eq.~\eqref{eq:2PAFrequency}. Except for the last term in~\eqref{eq:MComplete}, which does not depend on the integration frequency variable $\omega$, all the other terms in either~\eqref{eq:MComplete} or~\eqref{eq:Mb1} depend on $\omega$ through elementary factors of the form $(\omega-\omega_0)^{-1}$. The folding in~\eqref{eq:2PAFrequency}, therefore, can easily be carried out using Eq.~\eqref{eq:frequencyIntegral}.  In the case of the absorption of photon $1$ followed by that of photon $2$, the expression for the transition amplitude reads \begin{eqnarray}
\mathcal{A}_{\beta E,g}^{21}&=&
\mathcal{F}^{21}(\tau)\,e^{i\omega_2\tau}\bar{O}_{\beta\alpha}\mathcal{O}_{\alpha E, g}\,\frac{\epsilon_{E b}+i}{\epsilon_{E b}-i}\,\times\nonumber\\
\times&\Bigg\{&
\frac{\epsilon_{E a}+q_{\tilde{a}g}}{\epsilon_{E a}+i}w(z_E^{21})+(q_{\tilde{a}g}-i)w(z_{\tilde{E}_a}^{21})\times\nonumber\\
\times&\Bigg[&
\beta_a \frac{\epsilon_{Eb}+q_{\tilde{b}a}}{\epsilon_{E b}+i}-\frac{1}{\epsilon_{E a}+i}
+\frac{\delta_{ba}(q_{\tilde{a}b}-i)-\zeta_{ba}}{\epsilon_{E b}+i}
\Bigg]+\nonumber\\
+&&\sqrt{\frac{2}{\pi}}\frac{1}{\sigma_t}\frac{\mathcal{O}_{b,\alpha}}{V_{b,\beta}\bar{\mathcal{O}}_{\beta\alpha}(\epsilon_{E b}+i)}+\nonumber\\
+&&\sum_n\frac{\epsilon_{Eb}+q_{\tilde{b}n}}{\epsilon_{Eb}+i}\frac{\mathcal{O}_{\beta E, n}\mathcal{O}_{ng}}{\bar{O}_{\beta\alpha}\mathcal{O}_{\alpha E, g}} w(z_{E_n}^{21})
\,\Bigg\},\label{eq:A2Complete}
\end{eqnarray}
where we introduced the additional parameters
\begin{equation}
\delta_{ba}=\frac{\Gamma_a/2}{V_{b,\beta E}}\frac{\mathcal{O}_{b,\alpha}}{\bar{\mathcal{O}}_{\beta\alpha}},\quad
\zeta_{ba}=\frac{V_{a,\alpha}}{V_{b,\beta}}\frac{\mathcal{O}_{ba}}{\bar{\mathcal{O}}_{\beta\alpha}}.
\end{equation}
Equation~\eqref{eq:A2Complete}, which is one of the major results of this paper, depends on a minimal number of parameters for the radiative and non-radiative couplings between all the essential states involved in the dynamics, as well as the parameters of the pump and probe impinging pulses, including their time delay.  Once the parameters of the model are established, therefore, this formula is able to provide, at a negligible computational cost, full energy and time-delay resolved attosecond pump-probe photoelectron spectra in the presence of both an intermediate and a final resonance for arbitrary pairs of (weak) pulses. Furthermore, this result is trivially extended to the case of an arbitrary number of Gaussian pulses, to represent, e.g., the effect of an attosecond pulse train, as well as to an arbitrary number of isolated resonances either in the intermediate or in the final states.

It is now interesting to consider more in detail the case of no final resonances, for which the transition amplitude~\eqref{eq:A2Complete} simplifies to
\begin{eqnarray}
\mathcal{A}_{\beta E, g}^{21}&=&\mathcal{F}^{21}(\tau)\,e^{i\omega_2\tau}\bar{O}_{\beta\alpha}\mathcal{O}_{\alpha E, g}\,\times\nonumber\\
&\times&\Bigg[
\frac{\epsilon_{E a}+q_{\tilde{a}g}}{\epsilon_{E a}+i}w(z_E^{21})+\nonumber\\
&&+
\left(\beta_a-\frac{1}{\epsilon_{E a}+i}\right)(q_{\tilde{a}g}-i) w(z_{\tilde{E}_a}^{21})\,+\nonumber\\
&&+\sum_n\frac{\mathcal{O}_{\beta E, n}\mathcal{O}_{ng}}{\bar{O}_{\beta\alpha}(E)\mathcal{O}_{\alpha E, g}} w(z_{E_n}^{21})\,\Bigg],\label{eq:A2IntermediateResonance}
\end{eqnarray}
in relation to the simple two-photon transition matrix element~\eqref{eq:MIntermediateResonance} discussed earlier in this section.  In particular, we want to examine the effect of using finite pulses on the complex trajectory of the two-photon transition amplitude as a function of the central energy of the pump pulse. 
\begin{figure}[hbtp!]
\begin{center}
\includegraphics[width=\linewidth]{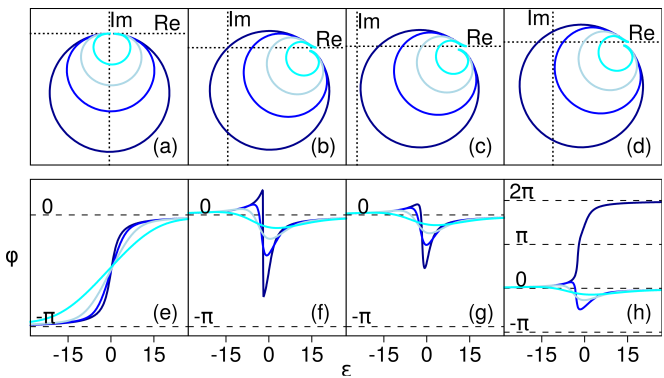}
\caption{(Color online) \label{fig:FoldedAmplitudes} Complex trajectories of resonant finite-pulse two-photon two-color absorption amplitudes (upper panels), and corresponding phase variation (lower panels) as the reduced intermediate energy detuning $\epsilon_{E_1a}$ increases from large negative ($-30$) to large positive values ($+30$). In each of the upper panels, increasingly shorter pulses ($\sigma_t\Gamma_a=\infty,\, 2,\, 1,\, 0.5$) give rise to progressively more contracted trajectories (traced with lighter color). Each column has a different set of $q$ and $\gamma$ parameters (see text for details).}
\end{center}
\end{figure}
Each of the upper panels in Fig.~\ref{fig:FoldedAmplitudes} shows the complex trajectories of the transition amplitude~\eqref{eq:A2IntermediateResonance}, with $\epsilon_{E_1a}\in[-30,30]$ and at four different pulse durations, $\sigma_t\Gamma_a=\infty,\, 2,\, 1,\, 0.5$, for a selected pair of resonance parameters:
a) $q_{\tilde{a}g}=20\gg1$, $\gamma_{a2}=0$,
b) $q_{\tilde{a}g}=1$, $\gamma_{a2}=0$,
c) $q_{\tilde{a}g}=1$, $\gamma_{a2}=0.2$,
d) $q_{\tilde{a}g}=1$, $\gamma_{a2}=-0.2$.
The amplitudes are normalised so to start at the same reference point on the real axis, which corresponds to the asymptotic background transition amplitude. The lower panels, Fig.~\ref{fig:FoldedAmplitudes}.e-h, show the transition amplitude phase as a function of the reduced pump detuning $\epsilon_{Ea}$. When the transition amplitude is dominated by the contribution of the intermediate autoionizing state ($q_{\tilde{a}g}\gg1$, Figs.~\ref{fig:FoldedAmplitudes}a,e), the observed phase excursion is always $\pi$. The shorter the pulse duration, the wider the step (Fig.~\ref{fig:FoldedAmplitudes}e).  If $q$ is finite but the intermediate autoionizing state is not radiatively coupled to the final continuum, the trajectory intercepts the origin, but only in the limit of long pulses, while, for short pulses, folding with the pulse spectra contracts the circular trajectory towards the asymptotic background value (Fig.~\ref{fig:FoldedAmplitudes}b). In particular, the phase loses its discontinuity, giving rise to a sigmoidal profile with no net phase change, with features that are progressively less pronounced as shorter pulses are employed (Fig.~\ref{fig:FoldedAmplitudes}f). The effect of finite pulses, therefore, is similar to that of a direct dipolar coupling between the bound state and the final continuum or, as we will see later in this section, to that of multiple intermediate channels. A similar dependence on pulse duration is observed for $\gamma_{a2}>0$ (Figs.~\ref{fig:FoldedAmplitudes}c,g). The complementary case of $\gamma_{a2}<0$ (if $\gamma_{a2}>0$ for probe absorption, $\gamma_{a2}<0$ for probe emission, and vice versa)  (Figs.~\ref{fig:FoldedAmplitudes}c,g) is more interesting because, in the long-pulse limit, the phase experiences a full $2\pi$ jump, transitioning to the continuous excursion with no net phase through a point, for a definite finite pulse duration $\sigma_t$, at which the phase has a discontinuous jump or, stated otherwise, at which the resonant two-photon transition amplitude exactly vanishes.

If the energy of the second photon is much larger than the natural width of the intermediate resonance, $\omega_2\gg \Gamma_a$,  Eq.~\eqref{eq:A2IntermediateResonance} can be further simplified. In fact, if the more stringent assumption $|\epsilon_{fa}| \gg q_a$ holds, we recover the expression given in Eq.(6) of~\cite{Jimenez2014}, 
\begin{equation}\label{eq:SimplifiedFormula}
\begin{split}
\mathcal{A}_{\beta E,g}^{21} \simeq &\,\mathcal{F}(\tau) e^{- i (\omega_2\tau + \phi_1+\phi_2)}\times\\
\times&\left[ w(z_E^{21}) + (\beta_{a}-\epsilon_{Ea}^{-1}) (q_{\tilde{a}g}-i)w(z_{\tilde{E}_a}^{21})\right],
\end{split}
\end{equation}
which was indeed justified in the context of helium photoionization in the region of the doubly excited states converging to the $N=2$ threshold.

\paragraph{Correspondence between intermediate-energy scan and final-energy resolved photoelectron spectrum.}
So far, when commenting the case of no final resonances, we have concentrated our attention on the phase of the resonant two-photon transition amplitude as a function of the central frequency $\omega_1$ of the pump pulse scanning the resonance, for a given value of the final-energy detuning $\delta$ from the nominal value $E_g+\omega_1+\omega_2$. Alternatively, one can keep $\omega_1$ constant and study the dependence of the transition amplitude on the final energy instead. In either cases, the variation of the amplitude is essentially dictated by the resonant argument $z_{\tilde{E}_a}^{21}$ of the Fadeeva function in~\eqref{eq:A2IntermediateResonance},
\begin{equation}\label{eq:Zequiv1}
z_{\tilde{E}_a}^{21} = \frac{\sigma_t}{\sqrt{2}} \left[
(\omega_{1}-\omega_{\tilde{a}g}) +\frac{\sigma_1^2}{\sigma^2}(E-E_g-\omega_1-\omega_2)-i\frac{\tau}{\sigma_t^2}\right],
\end{equation}
all the other terms in~\eqref{eq:A2IntermediateResonance} having, in comparison, only a weak dependence on $E$ and $\omega_1$.
From Eq.~\eqref{eq:Zequiv1}, the similarity between these two cases is evident: in the first case (scan over $\omega_1$), the second term in parenthesis is constant while the first increases linearly with $\omega_1$; in the second case (scan over $E$), the first term in parenthesis is constant while the second increases linearly with $E$.  In attosecond pump-probe experiments, furthermore, the pump pulse is oftentimes much shorter than the probe, and hence $\sigma_1^2/\sigma^2\simeq 1$. In these conditions, therefore, the two cases become essentially equivalent.

\paragraph{Monochromatic limit.} \label{par:monochromaticLimit}
It is instructive to ascertain that the formula for the finite-pulse resonant two-photon transition amplitude~\eqref{eq:A2IntermediateResonance} approaches the stationary expression~\eqref{eq:simpleMIntRes1} in the limit of long overlapping pulses, i.e., assuming pulse durations much longer than the resonance lifetime, $\sigma_t\Gamma_{a}\gg1$, and time delays negligible if compared to the duration of the pulses, $\tau\ll\sigma_t$. For $\sigma_t \rightarrow \infty$, the argument of the Faddeeva function, $z$, tends to $(+\infty,0)\in\mathbb{C}$, so one can use the first term in the asymptotic expansion of $w(z)$ restricted to the real axis, $w(x) \simeq i\pi^{-1/2}x^{-1}$, $x\in\mathbb{R}$~\cite{AbramowitzStegun}.
For $z^{21}_E\simeq\sigma_t/\sqrt{2}\left(E_i+\omega_1-\delta\sigma_1^2/\sigma^2-E\right)$, the Faddeeva function becomes $w(z^{21}_E) \simeq i\sqrt{2/\pi}\,\sigma_t^{-1} (E_i+\omega_1-\delta\sigma_1^2/\sigma^2-E)^{-1}$.
If we neglect the effect of intermediate bound states, we obtain
\begin{eqnarray}
\mathcal{A}_{\beta E,g}^{21}&\propto&
\frac{1}{E_g+\omega_1-\delta\sigma_1^2/\sigma^2-E}\frac{\epsilon_{E a}+q_{\tilde{a}g}}{\epsilon_{E a}+i}+\nonumber\\
&+&
\frac{1}{E_g+\omega_1-\delta\sigma_1^2/\sigma^2-\tilde{E}_a}\left(\beta_a-\frac{1}{\epsilon_{E a}+i}\right)(q_{\tilde{a}g}-i).
\nonumber
\end{eqnarray}
At the nominal energy of the transition ($\delta=0$), and using the energy-preserving condition $E=E_g+\omega_1+\omega_2$, the transition amplitude finally becomes
\begin{equation}
\mathcal{A}_{\beta E,g}^{21}\propto
\frac{\epsilon_{E_1a}+q^{(21)}_{\mathrm{eff}}}{\epsilon_{E_1a}+i},
\label{eq:LongPulseSimplified}
\end{equation}
as anticipated.

\paragraph{Case of non-overlapping pulses}\label{sec:IIICj}
In the presence of an intermediate resonance $|a\rangle$, instead of plummeting as soon as $|\tau|\geq\sigma_t$, as it was the case for non-resonant transitions (compare with Sec.~\ref{sec:IId}), the two-photon signal persists even for $\tau>\sigma_t$, decaying exponentially as $e^{-\tau/\tau_a}$ ($\tau_a=\Gamma_a^{-1}$).  To see this, let us consider the transition amplitude at zero final energy detuning, when $\tau\gg \sigma_t$, and the first photon absorption is right on resonance, $\omega_1=\mathrm{Re}[\omega_{ag}]$, 
\begin{equation}
z_a = \frac{i}{\sqrt{2}} \left(\frac{\sigma_t}{2\tau_a}-\frac{\tau}{\sigma_t}\right).
\end{equation} 
Then, the amplitude becomes proportional to
\begin{equation}
\mathcal{A}_{\beta E,g}^{21}\propto e^{-\tau^2/2\sigma_t^2}\,w(z_a)\simeq 2\, \exp\left( - \tau/2\tau_a\right),
\end{equation}
where we made the approximation $\text{erfc} (-iz) \sim 2 $ and we neglected the small term $\sigma_t^2/8\tau_a^2$. As expected, the transition amplitude decays exponentially with the time delay with half the lifetime of the resonant state.

Notice that for negative time delays the resonant signal still decays as the overlap of the pump and probe pulses (provided that the probe pulse is not itself in resonance with a transition from the ground state to a bound or autoionizing intermediate state). This latter circumstance illustrates how, in a time-resolved formulation, the time ordering of photon exchange in the transition matrix elements translates to an actual order in the two-photon transitions, when the two photons belong to non-overlapping pulses.
 
\subsection{Multichannel case}\label{sub:multichannel}
The results obtained thus far are valid only for the case of a single intermediate and a single final continuum channel.  As long as the intermediate and final resonances are isolated, and if all the coupling matrix elements involving the continua are smooth and slowly varying functions of the energy, however, generalisation to the case of an arbitrary number of intermediate and final continua is straightforward. As shown in Section 4 of the original Fano paper \cite{Fano1961}, the case of a bound state $|a\rangle$ coupled to several unperturbed continua $|\alpha \varepsilon\rangle$, $V_{\alpha a}=\langle\alpha \varepsilon |H|a\rangle$, can be reduced to that of the bound state $|a\rangle$ coupled to a \emph{single} ``resonant'' continuum $|R\,\varepsilon\rangle$, $V_{Ra}=\sqrt{\sum_{\alpha}\left|V_{\alpha a}\right|^2}=\langle R\, \varepsilon |H|a\rangle$, plus a set of fully decoupled residual featureless continua $|\alpha'\varepsilon\rangle$, $\langle \alpha'\varepsilon |H|a\rangle=0$, by means of a unitary transformation of the degenerate unperturbed continua, 
\begin{eqnarray}
&&|R\,\varepsilon\rangle = \sum_{\alpha} |\alpha\varepsilon\rangle U_{\alpha R},\quad |\alpha'\varepsilon\rangle = \sum_{\alpha} |\alpha\varepsilon\rangle U_{\alpha \alpha'},\\
&&U_{\alpha R}=V_{\alpha a}/V_{Ra}, \quad V_{Ra} \equiv \sqrt{\sum_{\alpha} |V_{\alpha a}|^2},\\
&&U^\dagger U = U U^\dagger =1.
\end{eqnarray}
 Furthermore, the residual decoupled continua $|\alpha'\varepsilon\rangle$ can be chosen so that only one of them, which we will call $|D\,\varepsilon\rangle$, is radiatively coupled to the ground state, $\mathcal{O}_{Dg}=\langle D\,\varepsilon|\mathcal{O}|g\rangle$, while the other continua are couple neither to the resonance nor radiatively to the ground state, and can therefore be entirely ignored. As a consequence, the transition amplitude $\mathcal{A}_{\beta E,g}$ to a single final continuum $\beta$ through a multichannel intermediate continuum can be reduced to the coherent sum of two amplitudes: one for a single resonant intermediate channel, $\mathcal{A}_{\beta E,R,g}$, and one for a single non-resonant intermediate channel, $\mathcal{A}_{\beta E,D,g}$,
\begin{equation}
\mathcal{A}_{\beta E,g}=\mathcal{A}_{\beta E,R,g}+\mathcal{A}_{\beta E,D,g}.
\end{equation} 
A similar reasoning applies to the final states, since even in that case it is possible to identify a single final resonant continuum. However, due to the presence of multiple intermediate states (two different continua, the bound states and the autoionizing state), more than two decoupled final continua can eventually be populated by means of a dipole transition. In any case, the final continua can be treated separately. If the final channel is not resolved, the individual contributions of all the final states to the total signal $\mathcal{P}_{E,g}$ must be added incoherently,
\begin{equation}
\mathcal{P}_{Eg}=\sum_{\beta} |\mathcal{A}_{\beta E,g}|^2.
\end{equation}
In other terms, the problem of several final channels can be treated as several problems of a single final channel (being it resonant or not).  In conclusion, if all the relevant couplings with the intermediate and final resonant and decoupled unperturbed channels are available, the multichannel problem can be treated as a combination of the amplitudes given earlier in this section.

Let us examine the case of one intermediate resonance in a multichannel continuum and no final resonances. According to the above and to Eq.~\eqref{eq:A2IntermediateResonance}, we can write
\begin{eqnarray}
\mathcal{A}_{\beta E,R, g}^{21}&=&\mathcal{F}^{21}(\tau)\,e^{i\omega_2\tau}\bar{O}_{\beta R}\mathcal{O}_{R g}\,\Big[
\frac{\epsilon_{E a}+q_{\tilde{a}g}}{\epsilon_{E a}+i}w(z_E^{21})+\nonumber\\
&+&
\left(\beta_a-\frac{1}{\epsilon_{E a}+i}\right)(q_{\tilde{a}g}-i) w(z_{\tilde{E}_a}^{21})\,\Bigg]+\label{eq:A2RIntermediateResonance}\\
&+&\mathcal{F}^{21}(\tau)\,e^{i\omega_2\tau}\sum_n\mathcal{O}_{\beta E, n}\mathcal{O}_{ng} w(z_{E_n}^{21}),\nonumber\\
\mathcal{A}_{\beta E, D,g}^{21}&=&\mathcal{F}^{21}(\tau)\,e^{i\omega_2\tau}\bar{O}_{\beta D}\mathcal{O}_{D g}\,w(z_E^{21}).
\label{eq:A2DIntermediateResonance}
\end{eqnarray}
When taking the sum of the resonant and decoupled amplitudes, the latter can be integrated in the first term in parenthesis of the former, giving rise to an effective complex $q$ parameter. The overall amplitude, however, cannot be assimilated to a single resonant transition amplitude by simply redefining the parameters involved. Thus, in principle, the presence of a decoupled channel qualitatively alters the finite-pulse resonant transition amplitude. In the long-pulse limit, however, the situation changes, as the total transition amplitude becomes proportional to [compare with Eq.~\eqref{eq:LongPulseSimplified}]
\begin{equation}\label{eq:multichanSimple1}
\mathcal{A}^{21}_{\beta E,g}\propto r_{DR}+\frac{\epsilon_{E_1a}+q^{(21)}_{\mathrm{eff}}}{\epsilon_{E_1a}+i},\quad r_{DR}= \frac{\bar{O}_{\beta D}\mathcal{O}_{D g}}{\bar{O}_{\beta R}\mathcal{O}_{R g}}.
\end{equation}
The constant term $r_{DR}$ expresses the strength of the dipolar coupling to the final continuum through the decoupled  intermediate continuum $|D\varepsilon\rangle$ relative to the one through the (unperturbed) resonant continuum $|R\varepsilon\rangle$. Equation~\eqref{eq:multichanSimple1} can be rewritten as
\begin{equation}\label{eq:multichanSimple2}
\mathcal{A}^{21}_{\beta E,g}\propto \frac{\epsilon_{E_1a}+q^{(21)}_{\mathrm{eff}'}}{\epsilon_{E_1a}+i},\quad 
q^{(21)}_{\mathrm{eff}'}= \frac{1+ir_{DR}}{1+r_{DR}}\,q^{(21)}_{\mathrm{eff}}.
\end{equation}
Therefore, in the long-pulse limit, the effect of multiple channels manifests itself as a simple modification of the effective complex $q$ parameter, exactly as it happened in the case of a finite dipolar coupling between the intermediate metastable state $|a\rangle$ and the final continuum $|\beta E\rangle$, $\gamma_{a2}\neq 0$. While in the latter case the modification of the effective $q$ was different for the absorption and for the emission of the second photon, however, in the multichannel case the variation of $q$ is identical for the two paths. In principle, therefore, it is still possible to disentangle the two effects by comparing these two transition amplitudes.

\subsection{Multiple intermediate and final resonances}
The total transition amplitude~\eqref{eq:A2Complete} can be generalised to the case of several intermediate and final isolated resonances by adding to the common background term the individual contribution from the intermediate and final states plus the residual contributions from all intermediate-final resonance pairs,
\begin{eqnarray}
\mathcal{A}_{\beta E,g}^{21}&=&\mathcal{F}^{21}(\tau)\,e^{i\omega_2\tau}\bar{O}_{\beta\alpha}\mathcal{O}_{\alpha E, g}\,\mathcal{W}_{\beta E,g}^{21},\nonumber\\
\mathcal{W}_{\beta E,g}^{21}&=&\mathcal{W}_{\beta E,g}^{21,\mathrm{bg}}+\sum_{a}\mathcal{W}_{\beta E,g}^{21,a}+\sum_{b}\mathcal{W}_{\beta E,g}^{21,b}+\sum_{ba}\mathcal{W}_{\beta E,g}^{21,ba},\nonumber
\end{eqnarray}
where
\begin{eqnarray}
\mathcal{W}_{\beta E,g}^{21,\mathrm{bg}}&=&w(z_E^{21})+\sum_n\frac{\mathcal{O}_{\beta E, n}\mathcal{O}_{ng}}{\bar{O}_{\beta\alpha}\mathcal{O}_{\alpha E, g}} w(z_{E_n}^{21}),\nonumber\\
\mathcal{W}_{\beta E,g}^{21,a}&=&(q_{\tilde{a}g}-i)\left[ \beta_a w(z_{\tilde{E}_a}^{21}) + \frac{w(z_E^{21})-w(z_{\tilde{E}_a}^{21})}{\epsilon_{E a}+i}  \right]
\nonumber\\
\mathcal{W}_{\beta E,g}^{21,b}&=&\frac{2i \,w(z_E^{21})}{\epsilon_{Eb}-i}+\sqrt{\frac{2}{\pi}}\frac{1}{\sigma_t}\frac{\mathcal{O}_{b\alpha}}{V_{b\beta}\bar{\mathcal{O}}_{\beta\alpha}}\frac{1}{\epsilon_{Eb}-i}+\nonumber\\
&+&\sum_n\frac{q_{\tilde{b}n}+i}{\epsilon_{Eb}-i}\frac{\mathcal{O}_{\beta E, n}\mathcal{O}_{ng}}{\bar{O}_{\beta\alpha}\mathcal{O}_{\alpha E, g}} w(z_{E_n}^{21})\nonumber\\
\mathcal{W}_{\beta E,g}^{21,ba}&=&
\frac{q_{\tilde{a}g}-i}{\epsilon_{E b}-i}\Bigg\{2i\frac{w(z_E^{21})-w(z_{\tilde{E}_a}^{21})}{\epsilon_{E a}+i}+\nonumber\\
&+&w(z_{\tilde{E}_a}^{21})\left[
2i\,\beta_a+q_{\tilde{b}a}-i+\delta_{ba}(q_{\tilde{a}b}-i)-\zeta_{ba}
\right]\Bigg\}.\nonumber
\end{eqnarray}
This approach has been employed to compute the spectrum of a sideband comprising the $2p^2$~{$^1$S$^e$} autoionizing state in the RABITT ionization of the helium atom from the ground state, when both the lower and the upper harmonics contributing to the resonant sideband were themselves in resonance with the $sp_{2,2}^+$ and the $sp_{2,3}^+$ {$^1$P$^o$} states, respectively~\cite{Jimenez2014}. 

\subsection{Multiphoton transitions}
The $n-$th order finite-pulse transition amplitude~\eqref{eq:2PAFrequency} is
\begin{eqnarray}
\mathcal{A}_{fg}^{(n)}&=&\frac{-i}{(2\pi)^{\frac{n}{2}-1}}\int\hspace{-3pt}\cdots\hspace{-3pt}\int \delta(\omega_{fg}-\Omega_n')\prod_{i=1}^n\left[\tilde{F}(\omega_i')d\omega_i'\right]\times\nonumber\\
&\times&\langle f|\mathcal{O}\prod_{i=1}^{n-1}\left[G_0^+( E_g+\Omega_{i}')\mathcal{O}\right]|g\rangle,
\end{eqnarray}
where $\Omega_i'=\sum_{j=1}^i\omega_j'$ and the factors in the last operator product are assumed to be ordered from right to left. As long as the \emph{on-shell} approximation is justified, the techniques employed in Sec.~\ref{sec:IIIB} for the two-photon transition matrix element can be used also to compute $n-$th order transition matrix element. Furthermore, if no more than one intermediate resonance contributes to the transition, the procedure followed in Sec.~\ref{sec:IIIC} can be subsequently applied to evaluate the folding with the field. A particularly relevant example that meets these conditions is the absorption of one pump photon $\omega_1$ followed by that of two probe photons $\omega_2$, when only the first intermediate continuum $|\psi_{\alpha \epsilon}\rangle$ is resonant while the second intermediate continuum $|\beta \epsilon\rangle$ and the last continuum $|\gamma \epsilon\rangle$ are not. In this case, the three-photon transition matrix element $\mathcal{M}_{\gamma E,g}^{(221)}$ is 
\begin{eqnarray}
\mathcal{M}_{\gamma E,g}^{(221)}&=&\langle \gamma E|\mathcal{O}G_0^+( E_g+\omega_1'+\omega_2')\mathcal{O}G_0^+( E_g+\omega_1')\mathcal{O}|g\rangle\simeq\nonumber\\
&=&\frac{\bar{\mathcal{O}}_{\gamma\beta}}{ E_g+\omega_1'+\omega_2'-E+i0^+}\mathcal{M}_{\beta E,g}^{(21)}(\omega_1').
\end{eqnarray}
The transition amplitude then becomes
\begin{eqnarray}
\mathcal{A}_{\gamma E,g}^{221}&=&-\frac{i\bar{\mathcal{O}}_{\gamma\beta}}{\sqrt{2\pi}}
\int d\omega_1'\mathcal{M}_{\beta E,g}^{(21)}(\omega_1')\tilde{F}_1(\omega_1')\times\nonumber\\
&\times&\int d\omega_2'\frac{\tilde{F}_2(\omega_{Eg}-\omega_1'-\omega_2')\tilde{F}_2(\omega_2')}{ E_g+\omega_1'+\omega_2'-E+i0^+}\simeq\nonumber\\
&\simeq&\frac{i\bar{\mathcal{O}}_{\gamma\beta}}{\omega_2}\hspace{-2pt}\int\hspace{-3pt} d\omega\mathcal{M}_{\beta E,g}^{(21)}(\omega)\tilde{F}_1(\omega)\tilde{F_2^2}(\omega_{Eg}-\omega),\label{eq:A3}
\end{eqnarray}
where in the last passage we have assumed that the spectrum of the absorption component of the probe pulse is localised around $\omega_2$ and we used the convolution theorem. This means that the three-photon amplitude $\mathcal{A}_{\gamma E,g}^{221}$ is equal, apart for a multiplicative factor, to the two-photon amplitude in which the frequency and spectral width of the probe field are larger by a factor of 2 and $\sqrt{2}$, respectively.

\section{Resonant RABITT spectrum of helium}\label{sec:Helium}

\begin{figure}[hbtp!]
\includegraphics[width=\linewidth]{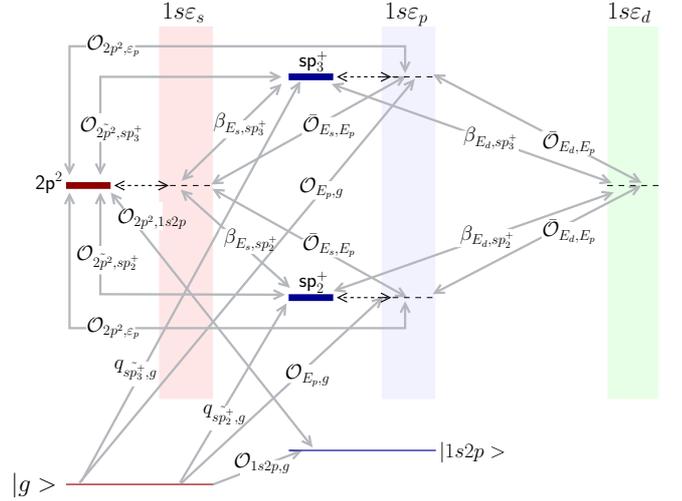}
\caption{\label{fig:EnergyScheme} (Color online).
Scheme of the essential states involved in the RABITT ionization of the helium atom in the region of the $N=2$ autoionizing states, together with the relevant radiative couplings between them that must be taken into account to reproduce the pump-probe photoelectron spectrum with the finite-pulse resonant two-photon model described in the text. The dashed lines indicate the non-radiative coupling between the resonant and the degenerate continuum states.}
\end{figure}

In this section we illustrate the capabilities of the finite-pulse two-photon resonant model by computing the RABITT photoionization spectrum of the helium atom from the $1s^2$~{$^1$S$^e$} ground state to the energy region between 30~eV and 40~eV above the first ionization threshold, which features the series of metastable doubly excited states that converge to the $N=2$ threshold.  Figure~\ref{fig:EnergyScheme} shows the energy levels of helium in the region of interest and illustrates schematically the radiative couplings that must be plugged in the model to reproduce the RABITT spectrum of the atom when the harmonics can be resonant with the first two {$^1$P$^o$} bright autoionizing states, and sidebands can populate the final $2p^2$ {$^1$S$^e$} state.
Helium is an ideal candidate to investigate atomic transitions through autoionizing states because it is amenable to an accurate \emph{ab initio} description in the presence of external light pulses. Furthermore, the ionization continuum of helium has been the subject of intense study for more than fifty years~\cite{Tanner2000}. In particular, the $N=2$ {$^1$P$^o$} and {$^1$S$^e$} series of autoionizing states have been investigated both experimentally~\cite{Madden1963,Domke1996,Schulz1996,Domke1991,Nagasono2007} and theoretically~\cite{Burke1963,Cooper1963a,Bhatia1973,Moccia1987,Macias1987,Macias1987b,Moccia1991,Sanchez1991,Lindroth1994,Burgers1995,Rost1997}, and many of their properties, such as positions, width, and $q$ parameter from the ground state, are well known. 

In the intermediate states of the model we included one {$^1$P$^o$} intermediate continuum, $1sE_p$, with the two isolated resonances $sp_2^+$ and $sp_3^+$~\cite{Cooper1963a}, and one intermediate bound state, $1s2p$, which contributes significantly to the excitation amplitude of the final $2p^2$ $^1$S$^e$ state, owing to the strong dipolar coupling between the $1s$ and $2p$ orbitals. In the final states of the model we included the $1sE_s$ $^1$S$^e$ continuum, featuring the $2p^2$ metastable state, and the $1sE_d$ {$^1$D$^e$} continuum. For the latter, we did not include any resonance, as the most relevant one, also with dominant configuration $2p^2$, lies very close in energy to the $sp_2^+$ state and hence it is not reached by any sideband within the chosen range of IR frequencies. The $sp_n^-$ and $2pnd$ states, as well as higher terms in the $sp_n^+$ series, which are all narrow and have a small dipole coupling with the ground state if compared with the $sp_{2/3}^+$ states, are not expected to affect significantly the sideband spectrum near or below the $2p^2$ $^1$S$^e$ state and were therefore not included in the model. 

The position, width and $q$ parameter from the ground state of the two intermediate resonances, as well as the background photoionization cross section, can be taken from the literature, where one can find also the position and width of the final $^1$S$^e$ state and the energy of the bound $1s2p$ state. Even with these data, there are still 14 independent parameters not reported in the literature that are in principle needed to apply the model: the two continuum-continuum couplings, $\bar{\mathcal{O}}_{1sE_\ell,1sE_p}$; the relative strength of the direct dipole coupling of the two intermediate autoionizing states with the two final continua, $\beta_{1sE_\ell,sp_{2}^+}$, $\beta_{1sE_\ell,sp_{3}^+}$  (4 parameters); the dipolar coupling to the two final continua through the intermediate $1s2p$ bound state, $\mathcal{O}_{1s \varepsilon_\ell,1s2p}\mathcal{O}_{1s2p,1s^2}$; the $q$ parameter for the excitation of the final resonance from the intermediate bound and the two intermediate autoionizing states, $q_{\tilde{2p}^2,1s2p}$, $q_{\tilde{2p}^2,sp_{2}^+}$, $q_{\tilde{2p}^2,sp_{3}^+}$, and, conversely, for the excitation of the $sp_{2/3}^+$ intermediate resonances from the $2p^2$ final metastable state, $q_{\tilde{sp}_{2}^+,2p^2}$, $q_{\tilde{sp}_{3}^+,2p^2}$; the dipole coupling between the final metastable state and the intermediate continuum, $\mathcal{O}_{2p^2,1s\varepsilon_p}$. The value of the residual parameters $\delta_{2p^2,sp_{2}^+}$, $\delta_{2p^2,sp_{3}^+}$, $\zeta_{2p^2,sp_{2,2}^+}$, and $\zeta_{2p^2,sp_{3}^+}$, can be determined from the previous ones with the additional assumption $\mathcal{O}_{2p^2,sp_{2/3}^+}\simeq\mathcal{O}_{\tilde{2p}^2,sp_{2/3}^+}$, which is justified by the strong dipole coupling between doubly excited states compared to that between an $N=2$ doubly excited state and a $1sE_\ell$ continuum (the latter optical transition, being itself a double excitation, is prohibited within the quasi-particle approximation). In the same spirit, we can assume $\mathcal{O}_{\tilde{sp}_{2/3}^+,2p^2}=\mathcal{O}_{sp_{2/3}^+,2p^2}$, with which $q_{\tilde{sp}_{2/3}^+,2p^2}$ become derived quantities, thus reducing to 12 the total number of independent parameters. Finally, we assume $q_{\tilde{2p}^2,1s2p}\gg 1$, so that the product $\mathcal{O}_{1s \varepsilon_\ell,1s2p}\mathcal{O}_{1s2p,1s^2}q_{\tilde{2p^2},1s2p}$ comes as a single parameter $\gamma_{\tilde{2p}^2\gets1s2p\gets1s^2}$, thus reducing the total number of parameters to 11. 

 In one of our previous works \cite{Jimenez2013}, we developed a soft-photon model for the non-resonant ionization of helium in which the continuum singlet states of the atom were approximated by the product of the $1s$  ground state of the He$^+$ parent-ion with free spherical waves,
\begin{equation}\label{eq:PlaneWaves}
\phi_{1s,E\ell m} = \frac{\alpha\beta-\beta\alpha}{\sqrt{2}}\frac{1+\mathcal{P}_{12}}{\sqrt{2}}\phi^{\mathrm{ion}}_{1s}(\vec{r_1})\sqrt{\frac{2k}{\pi}}j_{\ell}(kr_2)Y_{\ell m}(\hat{r_2}),
\end{equation} 
where $j_{\ell}$ are spherical Bessel functions, Y$_{\ell m}$ are spherical harmonics, and $E=k^2/2$, while the ground state was approximated by the $1s^2$ configuration, where, following Slater's prescription, the $1s$ orbital was an hydrogenic wavefunction with effective charge $Z=1.7$,
\begin{equation}
\phi^{\mathrm{He}}_{1s}(k)=\frac{2\sqrt{2}Z^{5/2}}{\pi[k^2+Z^2]^2}.
\end{equation}
With this model we were able to predict, with quantitative accuracy, the background distribution of the photoelectrons generated by the interaction of the atom with sequences of XUV-pump and IR-probe pulses, even in the presence of autoionizing resonances and for large IR intensities, provided that the whole spectrum (which occupied a limited energy region approximately 1~a.u. above the $1s$ threshold) was scaled by a constant factor $\mathcal{C}$, of the order of unity, which accounts for the known difference between the hydrogenic ionization cross section and the one predicted by the first Born approximation.  Since we are currently considering the same region of the photoelectron spectrum, it is justified to estimate the continuum-continuum couplings $\bar{\mathcal{O}}_{1sE_\ell,1sE_p}$ within the soft-photon approximation, further reducing the total number of free parameters to 9.

\begin{figure}[hbtp!]
\begin{center}
\includegraphics[width=\linewidth]{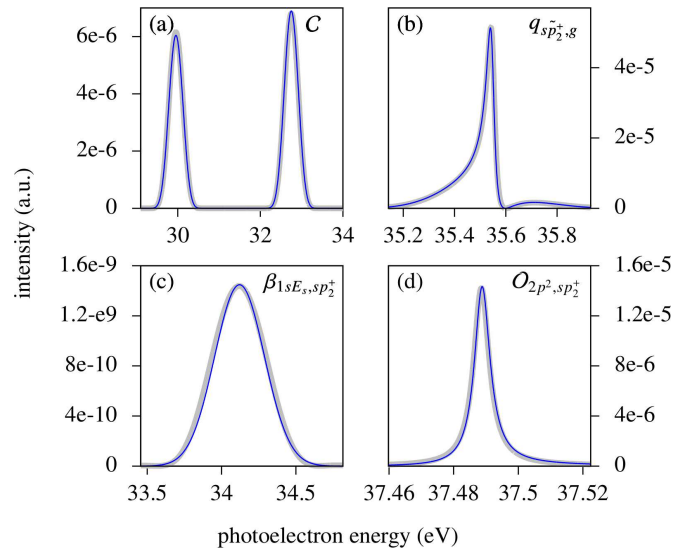}
\caption{\label{fig:CalibrationParameters} (Color online) Photoelectron spectra obtained \emph{ab initio} (thick gray solid line in the background) and with the model (thin black line in the foreground) for four different pulse sequences useful to calibrate the model parameters. a) non-resonant and b) resonant one-photon photoelectron spectrum obtained with using the XUV APT alone. The first spectrum (a) was used to determine the global scaling factor $\mathcal{C}$, while spectra like the one in (b) were used to confirm the parameters of the two $^1$P$^o$ autoionizing state. c) from the lower sideband of the $sp_2^+$ resonance generated with disjoint pump and probe pulses, we could estimate the resonance-continuum $\beta$ radiative couplings. d) from the resonant shape of the $sp_2^+$ state upper sideband, which strikes the $2p^2$ {$^1$S$^e$} state, we could determine the resonance-resonance dipolar coupling.}
\end{center}
\end{figure}
To determine the values of these parameters, and to subsequently verify the prediction of the model against reliable benchmarks, we carried out accurate \emph{ab initio} simulations based on the numerical solution of the time-dependent Schr\"odinger equation~\cite{Argenti2010,ArgentiATAS2012,Argenti2014,Jimenez2014} where the wave function for the two-active-electron system is represented in a B-spline~\cite{Bachau2001,Argenti2009} bipolar-spherical-harmonics~\cite{Varshalovich1988} close-coupling basis~\cite{Argenti2006,Argenti2007,Argenti2007a,Argenti2008b}, for both the time propagation and the asymptotic analysis of the single-ionization wave packet~\cite{Argenti2010,Lindroth2012,Argenti2013}, while the time-step propagation is carried out with a Krylov unitary approximation~\cite{Hochbruck1997} to a second-order split exponential time-evolution operator in velocity gauge, and implemented in a parallel code which makes use of the PETSc numerical library~\cite{petsc-web-page,petsc-user-ref,petsc-efficient}.
In the simulations, both the XUV APT and the IR probe have a duration (fwhm) of 6~fs, while their peak intensity is $I_{\XUV}=5$~GW/cm$^2$ and $I_{\IR}=10$~GW/cm$^2$, respectively. The individual attosecond pulses in the train have central energy $\omega_{\XUV} = 57$~eV and a duration of 250~as; consecutive pulses are separated by half the nominal IR period, $T_\IR=2\pi/\omega_\IR$. As we are particularly interested in the transition that, through the $sp_2^+$ and $sp_3^+$ ($^1P^o)$ DESs, populates the optically forbidden $2p^2$ ($^1S^e$) DES, we performed simulations for IR frequencies ranging from $\omega_{\IR} = 1.455$ eV to $\omega_{\IR} = 1.485$ eV. 

Using the XUV APT alone, we could determine the value of the scaling constant $\mathcal{C}$ to match the non-resonant background of the model to that of the \emph{ab initio} prediction (which agrees with the absolute value of the background photoionization cross section reported in the literature). With  $\mathcal{C} = 1.2$, the model and \emph{ab initio} backgrounds are in excellent agreement across an energy domain of $\sim 4\omega_{\IR}$ (see Fig.~\ref{fig:CalibrationParameters}a).  From the one-photon spectrum in the energy region where the $sp_2^+$ and $sp_3^+$ resonances are present, we determined position, width and $q$ parameter for the two autoionizing states: $\bar{E}_{sp_2^+}=-18.86$~eV, $\Gamma_{sp_2^+}=0.037$~eV, $q_{\tilde{sp_2^+}g}=-2.77$, $\bar{E}_{sp_3^+}=-15.34$~eV, $\Gamma_{sp_3^+}=0.0082$~eV, $q_{\tilde{sp_3^+}g}=-2.58$ (see Fig.~\ref{fig:CalibrationParameters}b), in agreement with the values reported in the literature~\cite{Rost1997}. 

We obtain the parameters $\beta_{1sE_\ell,sp_{2/3}^+}$ from the \emph{ab initio} background spectrum, resolved with respect to the orbital angular momentum $\ell$, of the two intermediate resonances for a time delay at which the pump and the probe pulses do not overlap since, as we already saw in the Sec.~\ref{sec:IIICj}, in this way the homogeneous contribution of the intermediate state vanishes. Notice that one could retrieve the same parameters also from the experiment by measuring the photoelectron spectrum at two different ejection angles. We determine both $\beta_{1sE_p,2p^2}$  and $\gamma_{\tilde{2p}^2\gets1s2p\gets 1s^2}$ by matching the parameters of the asymmetric resonant profile in the two-photon excitation of $2p^2$ with a pair of overlapping pump and probe pulses in which the harmonics are tuned out of resonance with respect to the intermediate autoionizing  states. Finally, to determine $q_{sp_{2/3}^+,2p^2}$, we look at the $2p^2$ resonant profile in the sideband of the \emph{ab initio} spectrum for non-overlapping APT and IR probe pulses, where alternatively the lower and the upper harmonics are in resonance with the $sp_{2}^+$ and the $sp_{3}^+$ state, respectively.

In Table~\ref{tab:parameters} we report the full list of the parameters that gave the best match with the \emph{ab initio} spectra discussed above.
\begin{table}[H]
\begin{center}
\caption{\label{tab:parameters}\footnotesize{ Radiative parameters for the two-photon resonant transitions model in helium. Atomics units are used.}}
\begin{tabular}{c||ccccc}
\hline\\[-2ex]
& $\bar{E}_{a}$ & $\Gamma_{a}$ & $q_{\tilde{a}}$ & $\beta_{1sE_s,a}$& $\beta_{1sE_d,a}$ \\ \hline\hline\\[-2ex]
a=sp$_2^+$\,\,&\,\,-0.693\,\,&\,\,1.37[-3]\,\,&\,\,-2.77\,\,&\,\, -0.003\,\, &\,\, -0.003\,\, \\ \hline\\[-2ex]
a=sp$_3^+$ &-0.564 & 3.01[-4] & -2.58 & -0.003 & -0.01 \\ \hline
\end{tabular}
\color{black}
\end{center}
\end{table}
\begin{table}[H]
\begin{center}
\begin{tabular}{c||cccccc}
\hline\\[-2ex]
& $\bar{E_b}$& $\Gamma_b$&$q_{b, sp_2^+}$ & $q_{b, sp_3^+}$& $\beta_{b, 1sE_p}$ &$\gamma_{\tilde{b}\gets1s2p\gets 1s^2} $ \\ \hline\\[-2ex]
b=2p$^2$&\,\,-0.622\,\, &\,\, 2.16[-4]\,\, &\,\, 153\,\, &\,\, 20\,\, &\,\, -0.003\,\, &\,\, 4255\\ \hline
\end{tabular}
\color{black}
\end{center}
\end{table}
Notice that almost all the parameters of the model were determined independently of each other by comparison with a minimal number of well defined selected numerical experiments. Alternatively, the parameters could have also been determined by comparing with actual time-unresolved experiments~\cite{Wuilleumier2006}.
Once these values are determined, the model can reproduce the photoelectron spectrum for several values of the IR frequency and pump-probe delay in the general case of partly overlapping pulses with multiple intermediate and one final resonant states, with no residual freedom to adjust the outcome.

\begin{figure*}[hbtp!]
\begin{center}
\includegraphics[width=\textwidth, scale=0.5]{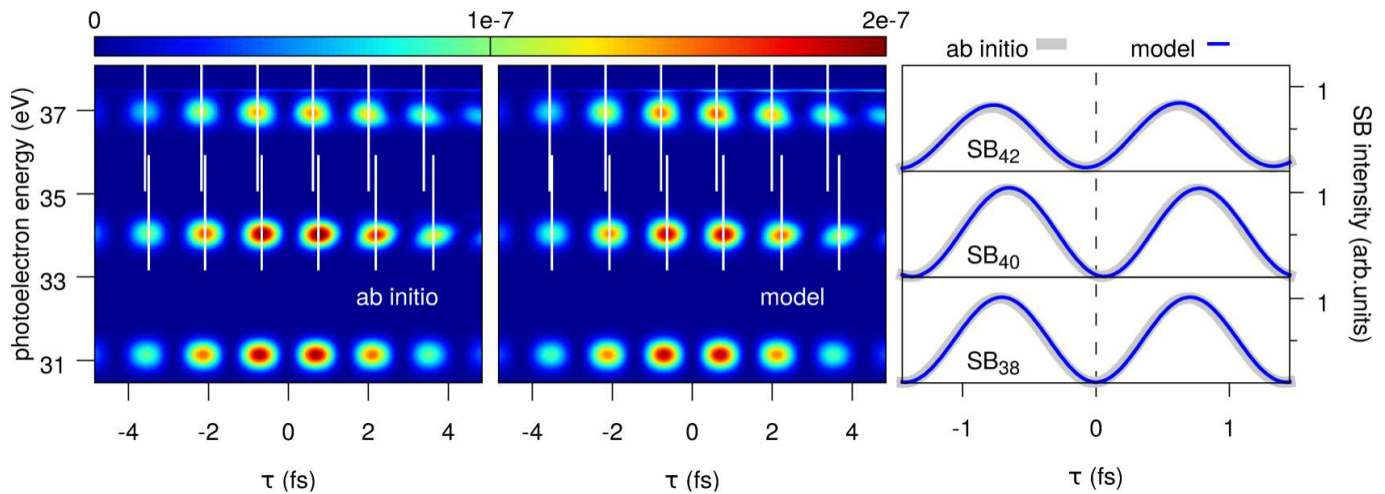}
\caption{\label{fig:FrequencyModulation} (Color online). Left panel: photoelectron spectrum as a function of the pump-probe time delay for sidebands SB$_{38}$, SB$_{40}$ and SB$_{42}$ of the driving frequency  $\omega_{\IR}=1.467$~eV. Right panel: energy-integrated photoelectron spectrum as a function of the pump-probe time delay for both model and \emph{ab initio} simulations. The presence of the sp$_2^+$ DES ($\sim 35.5$~eV), which is populated by the 41st harmonic (not shown) shifts sidebands SB$_{40}$ and SB$_{42}$ in opposite directions. Furthermore, due to the finiteness of the pulses used, the resonance induces a frequency modulation which can be seen by comparing the separation between the white lines that indicate the maxima of the sideband oscillations.}
\end{center}
\end{figure*}

The left panel of Fig.~\ref{fig:FrequencyModulation} shows both the \emph{ab initio} (left panel) and the model (central panel) prediction of the photoelectron spectrum as a function of the pump probe time delay for the sidebands SB$_{38-42}$ at a fixed IR frequency of $\hbar\omega_{\IR}=1.466$eV. 
Positive time delays indicate the XUV pulse train comes first. At the mentioned frequency, the 41st harmonic is resonant with the $sp_2^+$ ($^1P^o$) doubly excited state ($E_{sp_2^+} \approx 35.55$eV). The intermediate resonance induces a local phaseshift, in opposite directions, of the resonantly populated sidebands SB$_{40}$ and SB$_{42}$, an effect that is not present when a sideband is populated via non-resonant paths only, as in the case of SB$_{38}$. Model and \emph{ab initio} spectra look essentially the same. To illustrate qualitatively how the model compares with the \emph{ab initio} theory, on the right panel of Fig.~\ref{fig:FrequencyModulation} we show the two predictions for the energy integrated spectrum of the two sidebands as a function of the time delay, and they are indeed found to be in excellent agreement. The vertical white lines in the two first panels denote the maximum of the sideband signal and show how the absolute value of the local phaseshift $\delta\varphi(\omega_{\IR},\tau)$ of the two resonant sidebands increases with the time delay, 
\begin{equation}
I_{SB}(\tau) \propto \cos\left\{2\omega_{\IR}\tau + \delta\varphi(\omega_{\IR},\tau)\right\}.
\end{equation}
This means that the resonance introduces a modulation of the RABITT beating frequency itself. In this scenario, the concept of a global RABITT phase loses its meaning. In the cases we examined, however, the phase deviation is well approximated by a linear interpolation, $\delta\varphi(\omega_{\IR},\tau) \approx \delta\varphi_0(\omega_{\IR}) + \delta\omega(\omega_{\IR})\tau$, so
\begin{equation}
I_{SB} \propto \cos \left\{\left[2\omega_{\IR}+\delta\omega(\omega_{\IR})\right]\tau + \delta\varphi_0\right\}.
\end{equation}
The local phaseshift is affected by the apparent phaseshift at $\tau=0$ as well as by the modulation of the frequency, $\delta\omega(\omega_{\IR})$. As discussed in Sec.~\ref{sec:IIe}, the modulation of RABITT beating frequency appears even in absence of intermediate resonances, as a result of using finite pulses. The latter non-resonant effect, however, is always a shift towards the red, it does not depend much on the IR carrier frequency, and it rapidly disappears as longer pulses are employed. The resonant modulation of the sideband frequency, on the other hand, induces opposite shifts in the two resonant sidebands, it depends strongly on the detuning of the resonant harmonics from the intermediate autoionizing states, and it becomes sharper when longer pulses are used. The model prediction for the frequency modulation, obtained by Fourier analyzing the energy integrated sideband signal, is $\delta\omega = -0.073$~eV. The non-resonant redshift associated to the use of a 800~nm 6~fs probe pulse is comparatively large, $\delta\omega_{NR} = -0.038$~eV. Both the total and the non-resonant values are in agreement with those from the \emph{ab initio} calculation (the latter being estimated from the non-resonant sideband SB$_{38}$). By taking the difference between the total and the non-resonant values, the resonant contribution to the sideband frequency modulation, due to the $sp_2^+$ doubly excited state, is estimated as $\delta\omega_{sp_2^+} = -0.035$~eV, which corresponds to a change in the RABBIT period of 17 as.  

\begin{figure}[hbtp!]
\begin{center}
\includegraphics[width=\linewidth]{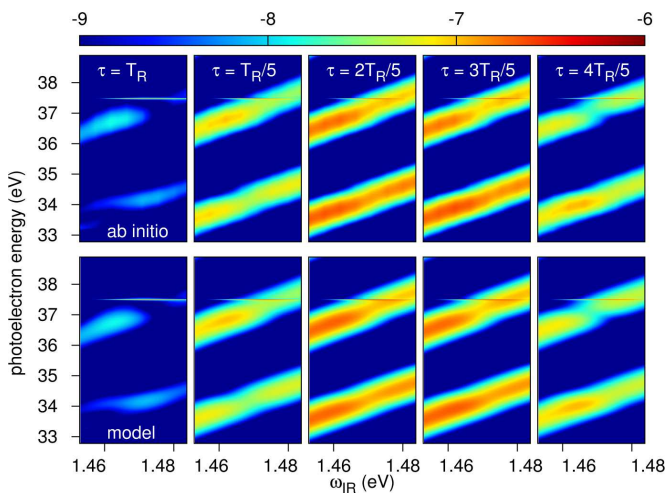}
\caption{\label{fig:PhaseShift} (Color online). Photoelectron spectrum as a function of the driving laser frequency for SB$_{40}$ and SB$_{42}$ at five different stages of the sideband oscillation ($\tau = T_{R}$ corresponds to the minimum). Upper panels show the \emph{ab initio} calculations and lower panels show the model results.}
\end{center}
\end{figure}

\begin{figure}[hbtp!]
\begin{center}
\includegraphics[width=\linewidth]{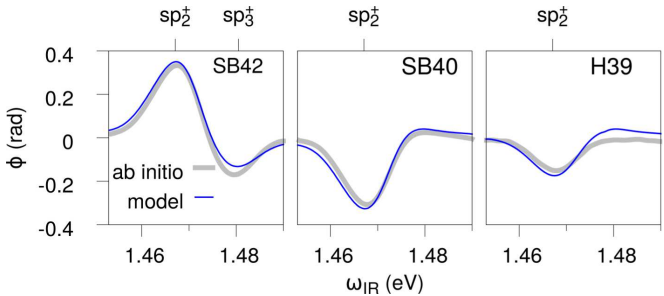}
\caption{\label{fig:ThirdOrder} (Color online) Apparent phaseshift of the integrated-signal beating of sidebands SB$_{42}$ (left panel), SB$_{40}$ (central panel), and harmonic H$_{41}$ (right panel), with respect to the non-resonant SB$_{38}$ sideband phase, as a function of the fundamental IR carrier frequency. In all three cases, the largest phase excursion occurs when the intermediate harmonic H$_{41}$ is resonant with the $sp_2^+$ DES. In SB$_{42}$, the effect of $sp_3^+$ through H$_{43}$ is visible as well. The resonant profile of H$_{41}$, which is comparable to the one of SB$_{40}$, results from the interference between the direct one-photon excitation amplitude of the continuum from the ground state with the three-photon amplitude, in which the resonant absorption of one pump photon from H$_{41}$ is followed by the stimulated emission of two IR probe photons.}
\end{center}
\end{figure}

Figure~(\ref{fig:PhaseShift}) shows the photoelectron spectrum of sidebands 40 and 42, as a function of the IR carrier frequency, for five different pump-probe time delays. 
The agreement between \emph{ab initio} and model, again, is excellent. In particular, the sideband resonantly populated from below (SB$_{42}$) shows a maximum to the left and a minimum to the right of the central resonance frequency ($\hbar \omega \approx$ 1.466 eV), while the opposite is true for the sideband that is populated from above.  This feature is responsible for the apparent phaseshifts for the two sidebands, which are shown in the first two panels of Fig.~\ref{fig:ThirdOrder} and were obtained by Fourier analyzing the spectrum in the time-delay interval $\tau\in[0,T_{IR}/2]$. The $sp_2^+$ and $sp_3^+$, populated by H41 and H43, respectively, give rise to resonant structures in the apparent phaseshift that are located at IR frequency close to the resonance condition of the each DES with the corresponding harmonics. As discussed in Sec.~\ref{sec:ResonantModel}, the overall phase excursion depends on the parameters of both the resonance and the pulses used. In the present case, the larger dipole matrix element of $sp_2^+$ with the ground state, compared with that of $sp_3^+$, makes the former dominate the shape of the profile, although the peak for the $sp_3^+$ DES can also be recognised. Finally, in the last panel of Fig.~\ref{fig:ThirdOrder} we show the phase of the beating of the H$_{39}$ integrated harmonic signal. In this case, the resonance profile arises from the interference between the direct one-photon ionization amplitude from the ground state and the \emph{three-photon} amplitude for the resonant absorption, from the ground state, of one XUV photon of the H$_{41}$ harmonic followed by the stimulated non-resonant emission of \emph{two} IR photons. 
As we discussed at the end of Sec.~\ref{sec:ResonantModel}, this latter amplitude can be easily computed with a straightforward multiphoton extension of the finite-pulse resonant model. The good agreement with the full-fledged \emph{ab initio} simulation for this process certifies that such extended model works.

\section{Conclusions} \label{sec:Conclusions}

In this work we have presented a new finite-pulse model for resonant two-photon transitions, which can be applied to simulate, at a negligible computational cost, attosecond pump-probe photoionization processes in atoms. The model, which extends the one presented in~\cite{Jimenez2014}, accounts for multiple intermediate and final channels, as well as the possible presence of multiple isolated resonances. Further generalisation to higher order transitions has been outlined.

We used the model to explain the physical origin of resonant phase profiles in two-photon ionization spectra as a function of the pump-resonance detuning. In particular, we showed that, if the intermediate states comprise a single continuum and a resonance not radiatively coupled to the final continuum states, the phase of the two-photon amplitude coincides with that of Fano one-photon transition, while in presence of multiple intermediate continua or of a finite radiative coupling between the intermediate resonance and the final continuum states, the phase experiences a continuous excursion with a net variation that can be either $0$ or $2\pi$. Furthermore, we showed that, when very short pulses are used, the beating frequency of the sidebands in the RABITT pump-probe scheme differs from the nominal $2\omega_\IR$ value. The results obtained with the model are found to be in quantitative agreement with virtually exact \emph{ab initio} simulations for the RABITT photoionization of helium in the region of the $N=2$ doubly excited states.

Even if benchmarked here against helium, the model is applicable to study time-resolved multiphoton resonant transitions in any atoms, molecules or solids susceptible of a description in terms of a finite number of free-particle channels and metastable states. Think, for example, of the radiative excitation of image-potential states on metal surfaces, which, on the one side, can decay by tunnelling to the conduction-band, and on the other side, can exchange a further photon and be liberated to either the metal or to the vacuum (photoemission channel)~\cite{Hofer1997,Muino2011}. 

When used as a phenomenological tool, the model can be employed to extrapolate, from time-resolved experiments with table-top laser apparatuses, the radiative-coupling strength between short-lived excited states, such as autoionizing states in heavier rare gases, which can be hard to obtain otherwise, either theoretically, due to the challenging role of electronic correlation, or experimentally, due to the need of coupling lasers to a synchrotron x-ray beamline~\cite{Wuilleumier2006}. Conversely, once the model is parametrised, it can be used as a computationally inexpensive alternative to the numerical integration of the TDSE. This is for example the case of photoemission studies conducted with the long, coherent and intense XUV pulses that became recently available at seeded XFEL facilities~\cite{Allaria2012a,Zitnik2014,Gauthier2015}.

\appendix

\section{Faddeeva function.}\label{app:Faddeeva}
In this appendix we derive the general analytical expression for the two-photon transition amplitude between an initial  state $|g\rangle$, with energy $E_g$, and a final state $|\beta E\rangle$, with energy $E$, due to the absorption/emission of a photon from a first Gaussian pulse $F_1$, centered in $t_1=0$, followed by the absorption/emission of a photon from a second Gaussian pulse $F_2$, centered in $t_2=t_1+\tau=\tau$,
\begin{equation}\label{eq:twoPhotonTransitionAmplitudeApp}
\mathcal{A}_{\beta E,g}^{21}=-i\int d\omega \,\tilde{F}_2(\omega_{Eg}-\omega;\tau)\tilde{F}_1(\omega)\mathcal{M}_{\beta E,g}(\omega),
\end{equation}
under the hypothesis, recurrent in the derivation of the model illustrated in Sec.~\ref{sec:ResonantModel}, that the two-photon matrix element $\mathcal{M}_{\beta E,g}(\omega)$ has an isolated simple pole at $\omega=E_a-E_g$, where $E_a\in\mathbb{C}$, $\mathrm{Im} E_a<0$, and that, to a very good approximation, $(\omega+E_g-E_a)\mathcal{M}_{\beta E, g}(\omega)$ is constant in the region where the product $\tilde{F}_2(\omega_{Eg}-\omega;\tau)\tilde{F}_1(\omega)$ does not vanish,
\begin{equation}
\mathcal{M}_{\beta E, g}(\omega)\simeq \frac{T_{\beta E,g}}{\omega-\omega_{ag}}.
\end{equation}
In the following, to consider all the possible cases at once, we will indicate both the absorption and the emission spectral components~\ref{eq:FourierTransformPulse} of the $n$-th Gaussian wavepacket~\eqref{eq:GaussianPulse} with the single expression
\begin{equation}\label{eq:App}
\tilde{A}_n(\omega)=\frac{A_n}{2\sigma_n}\,\mathrm{e}^{-i\varphi\sgn(\omega_n)} \mathrm{e}^{i \omega t_n} \mathrm{e}^{-\frac{(\omega-\omega_n)^2 }{2\sigma_n^2}},
\end{equation}
where the absorption/emission components are differentiated by attributing to $\omega_0$ a positive or negative sign, respectively. The two-photon transition amplitude~\eqref{eq:twoPhotonTransitionAmplitudeApp}, thus, becomes
\begin{eqnarray}\label{eq:twoPhotonTransitionAmplitudeGaussianApp}
\mathcal{A}_{\beta E,g}^{21}&=&-i\frac{A_{2}}{2\sigma_2}\,\mathrm{e}^{-i\varphi_2\sgn(\omega_2)}\frac{A_{1}}{2\sigma_1}\,\mathrm{e}^{-i\varphi_1\sgn(\omega_1)}\,\times\\
&\times&T_{\beta E,g}\,\int_{-\infty}^{\infty} \hspace{-12pt}d\omega \,\frac{e^{i(\omega_{Eg}-\omega)\tau}}{\omega-\omega_{ag}}e^{-\frac{(\omega_{Eg}-\omega-\omega_2)^2}{2\sigma_2^2}}e^{-\frac{(\omega-\omega_1)^2}{2\sigma_1^2}}.\nonumber
\end{eqnarray}
After some lengthy but straightforward algebraic passages, it is possible to cast the previous result in the following form
\begin{eqnarray}
\mathcal{A}_{\beta E,g}^{21}&=&-i\frac{A_{2}}{2\sigma_2}\,\mathrm{e}^{-i\varphi_2\sgn(\omega_2)}\frac{A_{1}}{2\sigma_1}\,\mathrm{e}^{-i\varphi_1\sgn(\omega_1)}\,\times\label{eq:twoPhotonTransitionAmplitudeGaussianApp2}\\
&\times&
\exp\left(-\frac{\delta^2}{2\sigma^2}-\frac{\tau^2}{2\sigma_t^2}-i\frac{\sigma_2}{\sigma_1}\frac{\tau}{\sigma_t}\frac{\delta}{\sigma}+i\omega_2\tau\right)\,\times\nonumber\\
&\times&T_{\beta E,g}\,\int_{-\infty}^{\infty} \hspace{-12pt}d\omega \,
\frac{\exp\left[-\frac{1}{2}\left(\sigma_t\omega+\frac{\sigma_1}{\sigma_2}\frac{\delta}{\sigma}+i\frac{\tau}{\sigma_t}\right)^2\right]}{\omega_1+\omega-\omega_{ag}},\nonumber
\end{eqnarray}
where we have introduced a convoluted spectral width $\sigma=\sqrt{\sigma_1^2+\sigma_2^2}$ and temporal width $\sigma_t=\sqrt{\sigma_1^{-2}+\sigma_2^{-2}}$ (notice that $\sigma=\sigma_1\sigma_2\sigma_t$), as well as the nominal detuning $\delta=E_g+\omega_1+\omega_2-E$.
By performing the change of variable $x=-\frac{1}{\sqrt{2}}(\sigma_t\omega+\frac{\sigma_1}{\sigma_2}\frac{\delta}{\sigma}+i\frac{\tau}{\sigma_t})$, the integral in Eq.~\eqref{eq:twoPhotonTransitionAmplitudeGaussianApp2} can be expressed in terms of the Faddeeva special function $w(z)=e^{-z^2}\mathrm{erfc}(-iz)$, which, in the upper half of the complex plane, admits the following integral representation (see \S7.1.3-4 in~\cite{AbramowitzStegun}),
\begin{equation}\label{eq:FaddeevaIntegral}
w(z)=\frac{i}{\pi} \int_{-\infty}^{+\infty} \frac{\mathrm{e}^{-t^2}}{z-t} dt, \qquad \text{Im}[z]>0.
\end{equation}
Indeed, by introducing the dimensionless complex variable $z_a^{21}$,
\begin{equation}
z_a^{21}=\frac{\sigma_t}{\sqrt{2}}\left[\left(\omega_1-\frac{\sigma_1^2}{\sigma^2}\delta-i\frac{\tau}{\sigma_t^2}
\right)-\omega_{ai}\right],
\end{equation}
the integral in Eq.~\eqref{eq:twoPhotonTransitionAmplitudeGaussianApp2} can be expressed as
\begin{equation}
\int_{-\infty}^{\infty} \hspace{-12pt}d\omega \,
\frac{\exp\left[-\frac{1}{2}\left(\sigma_t\omega+\frac{\sigma_1}{\sigma_2}\frac{\delta}{\sigma}+i\frac{\tau}{\sigma_t}\right)^2\right]}{\omega_1+\omega-\omega_{ai}}=-i\pi w(z_a).
\end{equation}
Notice that to establish the correspondence between the integral in Eq.~\eqref{eq:twoPhotonTransitionAmplitudeGaussianApp2} and the r.h.s. of~\eqref{eq:FaddeevaIntegral}, one must continuously deform the integration path from the initial real axis to a final re-defined real axis without crossing the pole, which requires $\tau<\sigma_t^2\,\mathrm{Im}E_a$. 
Once the integral is written in terms of the Faddeeva function, however, the expression is valid for any value of the time delay, since the Faddeeva function is defined on the whole complex plane by analytic continuation. The time-ordered two-photon transition amplitude finally becomes
\begin{eqnarray}\label{eq:twoPhotonTransitionAmplitudeGaussianApp3}
\mathcal{A}_{\beta E,g}^{21}&=&-\pi\frac{A_{1}A_{2}}{4\sigma_1\sigma_2}\,\mathrm{e}^{-i\varphi_2\sgn(\omega_2)}\mathrm{e}^{-i\varphi_1\sgn(\omega_1)}\,T_{\beta E,g}\,e^{i\omega_2\tau}\times\nonumber\\
&\times&
\exp\left[-\frac{1}{2}\left(\frac{\delta^2}{\sigma^2}+\frac{\tau^2}{\sigma_t^2}+2i\frac{\sigma_2}{\sigma_1}\frac{\delta}{\sigma}\frac{\tau}{\sigma_t}\right)\right]\,w(z_a^{21}).\nonumber
\end{eqnarray}

\begin{acknowledgments}
We thank Richard Ta\"ieb, Alfred Maquet, Jeremie Caillat, Pascal Sali\`eres, Eva Lindroth, Anne L'Huillier, Marcus Dahlstr\"om, and  Anatoli Kheifets for useful discussions. We thank Carlos Marante for providing us the original data for Fig.~\ref{fig:ContContTransAmp}. We acknowledge support from the European Research Council under the European Union's Seventh Framework Programme (FP7/2007-2013)/ERC grant XCHEM 290853, the MINECO project no. FIS2013-42002-R, the ERA-Chemistry Project PIM2010EEC-00751, the European grant MC-ITN CORINF and the European COST Action XLIC CM1204. Calculations were performed at the Centro de Computaci{\'o}n Cient{\'i}fica of the Universidad Aut{\'o}noma de Madrid and the Barcelona Supercomputer Centre. 
\end{acknowledgments}


%

\end{document}